\documentclass[10pt,a4paper]{article}

\usepackage[utf8]{inputenc}
\usepackage {amsmath, amssymb, mathrsfs, graphics, setspace}
\usepackage{braket}
\usepackage{multirow}
\usepackage{amsthm}
\usepackage{bbold}
\usepackage{cancel}
\usepackage{slashed}
\usepackage{color}
\usepackage{bbm}
\usepackage{amsfonts}
\usepackage{ytableau}
\usepackage{graphicx}
\usepackage{cite}
\usepackage{nicefrac}
\usepackage[affil-it]{authblk}
\usepackage{setspace}
\usepackage{placeins}
\usepackage[linktocpage=true]{hyperref}
\usepackage{xcolor}
\hypersetup{
    colorlinks,
    linkcolor={blue!90!black},
    citecolor={blue!90!black},
    urlcolor={blue!90!black}
}

\newcommand{\Op}{\mathcal{O}}
\newcommand{\idd}{\mathbb{1}}
\newcommand{\p}{^{\prime}}
\newcommand{\at}{\bigg\rvert}
\newcommand{\vet}{\boldsymbol}
\newcommand{\oneh}{\frac{1}{2}}
\newcommand{\de}{\partial}

\newcommand{\smatrix}{\begin{pmatrix}}
\newcommand{\cmatrix}{\end{pmatrix}}

\newcommand{\da}{^\dagger}
\DeclareMathOperator{\Tr}{Tr}

\DeclareMathOperator{\ide}{\mathbbm{1}}

\title{\vspace{-3.0cm}Multi-Representation Dynamics of $SU(4)$ Composite Higgs Models: Chiral Limit and Spectral
Reconstructions\vspace{0.5cm}}

\author[a]{Luigi Del Debbio\thanks{\href{mailto:luigi.del.debbio@ed.ac.uk}{luigi.del.debbio@ed.ac.uk}}}
\author[a]{Alessandro Lupo\thanks{\href{mailto:alessandro.lupo@ed.ac.uk}{alessandro.lupo@ed.ac.uk}}}
\author[b]{Marco Panero\thanks{\href{mailto:marco.panero@unito.it}{marco.panero@unito.it}}}
\author[c]{Nazario Tantalo\thanks{\href{mailto:nazario.tantalo@roma2.infn.it}{nazario.tantalo@roma2.infn.it}}}

\affil[a]{Higgs Centre for Theoretical Physics, School of Physics \& Astronomy, The University of Edinburgh, Peter Guthrie Tait Road, Edinburgh EH9 3FD, United Kingdom\vspace{0.8cm}}

\affil[b]{Department of Physics, University of Turin \& INFN, Turin\\
Via Pietro Giuria 1, I-20125 Turin, Italy\vspace{0.8cm}}

\affil[c]{University and INFN of Roma Tor Vergata\\
	Via della Ricerca Scientifica 1, I-00133, Rome, Italy}

\date{}
\usepackage{lineno}

\begin{document}
\maketitle
\begin{abstract}
\noindent We present a lattice study of the $SU(4)$ gauge theory with two Dirac fermions in the fundamental and two in the two-index antisymmetric representation, a model close to a theory of partial compositeness~\cite{Ferretti_2014su4}. Focus of this work are the methodologies behind the computation of the spectrum and the extrapolation of the chiral point for a theory with matter in multiple representations. While being still technical, this study provides important steps towards a non-perturbative understanding of the spectrum of theories of partial compositeness, which present a richer dynamics compared to single-representation theories. The multi-representation features are studied first in perturbation theory, and then non-perturbatively by adopting a dual outlook on lattice data through a joint analysis of time-momentum correlation functions and smeared spectral densities.
\end{abstract}
\newpage
\tableofcontents
\newpage
\section{Introduction}
The Standard Model (SM) of particle physics describes the strong and electro-weak (EW) interactions with remarkable accuracy, and no clear deviation from its predictions has been observed. There are however open problems pointing towards the idea that the SM effectively describes Nature only up to a cutoff energy scale $\Lambda_{SM}$ located at least in the TeV range. One of these issues, known as “Naturalness problem”, lies in the Higgs sector, where quantum corrections are expected to push the mass of the Higgs boson towards $\Lambda_{SM}$. The experimental value \cite{Aad_2012, Chatrchyan_2012}, however, lies well within the EW range. This value could only be understood within the SM by relying on fine-tuned cancellations, which are considered unsatisfactory from a theoretical perspective. The solutions that have been proposed to tackle this fine-tuning issue have been generating a vast and fertile literature, including ideas such as supersymmetry and technicolor. Another popular solution, and focus of this work, is the “composite Higgs” scenario \cite{Dugan:1984hq} where the Higgs boson's mass is explained in terms of Goldstone dynamics. A gauge theory is postulated to describe a new strongly-interacting sector and its dynamics. Depending on the fermionic content of the theory, a global flavor symmetry is realised: if the fermions condensate, the spontaneous breaking of such symmetry generates Goldstone bosons, among which there would be the Higgs doublet, the lightest state of the new sector. 

A number of models are compatible with the composite scenario. A set of minimal candidate gauge theories have been identified in Ref.~\cite{Ferretti_2014}. Such theories exhibit a doublet compatible with the SM Higgs boson in the low energy theory, together with other interesting phenomenological features such as asymptotic freedom and the presence of a custodial symmetry. Remarkably, these theories yield a composite state with the same SM quantum numbers as the heavy quarks. This crucial property could clarify the hierarchical structure of the quark masses: if the composite partner of the top quark has a large enough anomalous dimension, the mass hierarchy arises naturally. This idea goes under the name of partial compositeness~\cite{KAPLAN1991259} and it has been the subject of several lattice studies in recent years~\cite{Ayyar:2018glg, Ayyar:2018zuk, Ayyar:2017qdf, Cossu_2019, Bennett:2022yfa, Lupo:2021nzv}. The phenomenology of theories of partial compositeness is rich and can differ from Quantum Chromodynamics (QCD) in many aspects. It is common for these theories to involve fermions transforming in multiple representations of the gauge group. While QCD and its fermions in the fundamental representation have been extensively studied for several decades leading to tremendous improvement in the field, theories with multiple representations are still at an earlier stage, despite recent notable progresses~\cite{Ayyar:2018glg, Ayyar:2018zuk, Ayyar:2017qdf, Cossu_2019, Bennett:2022yfa}. A promising model of partial compositeness has been proposed by Ferretti~\cite{Ferretti_2014su4}. Its UV completion is an $SU(4)$ gauge theory featuring three Dirac fermions in the fundamental (Fund) and antifundamental representation, and five Majorana fermions in the two-index antisymmetric (2AS) representation of the gauge group. In this work, following the work presented in Refs.~\cite{Cossu_2019, Ayyar:2018glg, Ayyar:2018zuk, Ayyar:2017qdf}, we perform a lattice study of a simplified version of the Ferretti model, containing two Dirac fermions in the fundamental and antifundamental  and two Dirac fermions in the AS representation of the gauge group $SU(4)$. This model does not form a bound state compatible with the Higgs boson in the low-energy limit, but it is expected to maintain some of its non-perturbative features, and it therefore represents a solid starting point towards a better understanding of this class of theories. 

Our investigations focus on the methodologies required to address theories with matter in multiple representations. Although lattice simulations provide a flexible framework which has been well established over many decades, theories of partial compositeness present challenging features, and a dedicated study is an important step in order to ultimately make contact with their phenomenology. We will show that a richer dynamics, a distinguishing mark of these theories, complicates tasks such as the computation of the spectrum or the extrapolation to the chiral limit. For the model under consideration, we have used numerical simulations and perturbative calculations in order to clarify such dynamics. In particular, we have generalised previous perturbative results to the case of multi-representation theories in order to predict the critical mass of Wilson fermions. We have generated gauge configurations at different fermionic masses, enabling the analysis of the theory at the chiral point. For the first time, we performed a comparison of mesonic masses estimated from correlation functions and spectral densities: these two quantities, related by a Laplace transform, provide a dual outlook on non-perturbative data obtained on the lattice. The computation of spectral densities from lattice correlators has been the subject of several studies leading to interesting progress in the field \cite{Hansen:2017mnd, Hansen_2019, Bulava:2019kbi,  Bulava:2021fre, Gambino:2022dvu, Bruno:2020kyl, Bailas:2020qmv} and it is here performed for the first time in the context of BSM physics, where the lack of phenomenological inputs and a more sophisticated dynamics call for new computational strategies. 

The rest of this work is structured as follows: Sec.~\ref{sec:model} examines the model introduced by Ferretti and its simplified version with less matter content. Sec.~\ref{sec:pt} contains perturbative results for the critical mass of Wilson fermion, with a focus on the multi-representation dynamics. The lattice setup is outlined in Sec.~\ref{sec:lattice_setup}, and the observables targeted in our numerical simulations are described in Sec.~\ref{sec:observables}. In Sec.~\ref{sec:chiral_limit} we present results for the extrapolations of our data to the chiral point. Sec.~\ref{sec:backus-gilbert_1} is dedicated to the extraction of spectral densities from lattice correlators. Sec.~\ref{sec::measurements} contains a discussion on features of the spectrum of multi-representation theories. Sec.~\ref{sec:spectre} concludes the discussion with a description of methodology and results of fits of spectral densities, which are compared with the study of lattice correlators in the time-momentum representation.

\section{Models}
\label{sec:model}

\subsection{The Ferretti model}
We briefly recall the features of the model introduced
by Ferretti in Ref.~\cite{Ferretti_2014su4} following the conventions of~\cite{Cossu_2019}. Its UV completion is described by the gauge group $\mathcal{G}=SU(4)$. The gauge field is coupled to three Dirac fermions that we express in terms of Weyl doublets $\chi^a_m$, $\bar{\chi}^a_m$ respectively in the fundamental and antifundamental representation of the gauge group, together with five Majorana fermions $\psi^I_{mn}$ in the 2AS representation, which is real and dimension 6. The indices $a=1,2,3$ and $I=1,\dots 5$ are flavor indices, while $m,n=1,2,3,4$ denote the color. This matter content induces a global symmetry described by the group $G$
\begin{equation}
    G = SU(5) \times SU(3) \times SU(3)\p \times U(1)_X \times U(1)\p \; .
\end{equation}
  The charges of the fermions with respect to the flavor group are described in Table \ref{tab:globalflavor}.
\begin{table}[h!]
  	\centering
  	\begin{tabular}{||c c c c c c||} 
  		\hline
  		$\,$ & $SU(5)$ & $SU(3)$ & $SU(3)'$ & $U(1)_X $ & $U(1)' $ \\ [0.5ex] 
  		\hline\hline
  		$\psi$     &$\boldsymbol{5}$ & 	$\boldsymbol{1}$ & 	$\boldsymbol{1}$ & 0 & -1 \\  [0.5ex] 
  		\hline 
  		$\chi$     &$\boldsymbol{1}$ & 	$\boldsymbol{3}$ & 	$\boldsymbol{1}$ & -1/3 & 5/3 \\  [1ex] 
  		\hline 
  		$\tilde{\chi}$ &$\boldsymbol{1}$ & 	$\boldsymbol{1}$ & 	$\boldsymbol{\bar{3}}$ & 1/3 & 5/3 \\   
  		\hline
  	\end{tabular}
  	\caption{Flavor charges of the fermions in the Ferretti model.}
  	\label{tab:globalflavor}
  \end{table}
  
Neglecting couplings with the SM fermions, $G$ is an exact symmetry. Spontaneous symmetry breaking happens once the bilinears for both representations acquire a non-vanishing expectation value, leaving the unbroken subgroup $H$. The quotient group determining the low-energy dynamics is
  \begin{equation}
  \frac{G}{H} = \frac{SU(5)\times SU(3) \times SU(3)' \times U(1)_X \times U(1)' }{SO(5) \times SU(3)_c \times U(1)_X } \; .
  \end{equation}
This symmetry breaking pattern is interesting for several reasons. Given that $SO(5) \supset SU(2)\times SU(2)$, the pattern is compatible with the requirement of custodial symmetry $H \supset G_{\textrm{cust}} \supset G_{\textrm{SM}}$, with $G_{\textrm{cust}} = SU(3)_c \times SU(2)_L \times SU(2)_R \times U(1)_X$ and $G_{\textrm{SM}}$ being the SM gauge group $SU(3)_c \times SU(2)_L \times U(1)_Y$. The unbroken group $SU(3)_c$, related to the fundamental sector, is responsible for the strong interaction of QCD once it is gauged. The unbroken group related to the 2AS fermions, $SO(5)$, contains the EW group $SU(2)_L \times U(1)_Y$. In fact, $SO(5) \supset SO(4) \simeq SU(2)_L \times SU(2)_R$. We then define an $U(1)_R$ as the subgroup of $SU(2)_R$ generated by the generator of isospin rotation $T_R^{(3)}$: the correct hypercharges $Y$ are then obtained by $Y=T_R^{(3)}+X$, $X$ being the charge under $U(1)_X$. The quotient $SU(5)/SO(5)$ is therefore the relevant one for EW symmetry breaking: by writing its 14 Goldstone bosons in terms of SM charges,
  \begin{equation}
  \boldsymbol{14} \rightarrow \boldsymbol{1}_0 + \boldsymbol{2}_{\pm 1/2} + \boldsymbol{3}_0 \pm \boldsymbol{3}_{\pm 1} \equiv (\eta,H,\phi_0,\phi_\pm) \; ,
  \end{equation}
 we identify an $SU(2)$ doublet $\boldsymbol{2}_{\pm 1/2}$, $H$, that is compatible with the Higgs boson.
 
Turning to the composite partner for the top quark, this is introduced as a Dirac fermion $\Psi$ \cite{Ferretti_2014su4} in the low energy theory that has charges $(\vet{5},\vet{3})_{2/3}$ with respect to the unbroken subgroup $H$. States with these quantum numbers, relevant for partial compositeness, are obtained in this theory by color singlet combinations of fermions in different representations. Such baryonic content is therefore typical of theories with multi-representation matter.

\subsection{Two-flavor Ferretti model}
In this work we will focus on a simplified version of the Ferretti model. We will consider two Dirac fermions in the fundamental and two Dirac fermions in the 2AS representation of the gauge group $SU(4)$, a model that has been already studied in \cite{Ayyar:2018glg, Ayyar:2018zuk, Ayyar:2017qdf, Cossu_2019, Lupo:2021nzv}. While retaining the multi-representation dynamics and some non-perturbative features, this choice changes the global symmetries of the theory. 

It is important to understand the discrete symmetries of each sector in order to give the correct interpretation to the lattice data. Isospin, in particular, is useful in classifying scattering processes. The isospin group in the fundamental sector is the well known $SU(2)$. Since the symmetry breaking pattern is the same as in massless two-flavor QCD, characterised by the quotient group
\begin{equation}
     \frac{SU(2)_\mathrm{L} \times SU(2)_\mathrm{R}}{SU(2)_\mathrm{V}} \; ,
\end{equation}
and does not require to be discussed here. For completeness, we only mention that the three Goldstone bosons $\pi_1, \pi_2, \pi_3$ arising from these cosets can be labelled with eigenvalues of the azimuthal component of the isospin, $\pm 1, 0$:
\begin{equation}
    \begin{split}
        & \pi_\mathrm{+} = \frac{\pi_1 + i \pi_2}{\sqrt{2}} \; ,\\
        & \pi_\mathrm{-} = \frac{- \pi_1 + i \pi_2}{\sqrt{2}} \; , \\
        & \pi_\mathrm{0} = -\pi_3 \; \; .
    \end{split}
\end{equation}
The multiplet $\left(\pi_+, \pi_0,\pi_-\right)$ has eigenvalue $-1$ under the $G$-parity defined by combining charge conjugation $\mathcal{C}$ with an $\mathrm{SU(2)}$-isospin rotation,
\begin{equation}
    G = \exp \left(i \pi \tau_2 \right) \; \mathcal{C} \; ,
\end{equation}
$\tau_i$ being $SU(2)$ generators. In the 2AS sector, instead, the symmetry breaking pattern yields the cosets
\begin{equation}
    \frac{SU(4)}{SO(4)} \; .
\end{equation}
The generators $T_i$ of $SU(4)$ can be found in Appendix \ref{app:groups}. These cosets are characterised by 9 Goldstone bosons $\Pi_i$, $i=1,\dots 9$, that we represent exponentially as
\begin{equation}
    U = \exp (i \hat{T}_i \Pi_i) \; , 
\end{equation}
where $\hat{T}_i$, $i=1,\dots 9$ are the broken generators of $SU(4)$. Under an isospin transformation,
\begin{equation}
    U \rightarrow h U h\da \; , \;\;\;\;\; h=\exp(i \omega_n X_n) \; , \;\;\;\;\; n=1,\dots 6 \;,
\end{equation}
where $X_n$ are the unbroken generators of $SU(4)$ generating $SO(4)$, represented as $9\times 9$ matrices. We give such a representation in Appendix \ref{sec:app:iso}. The maximum set of commuting generators here is two, therefore we choose to diagonalise $X_1$ and $X_6$. The Goldstone bosons in the isospin basis can be then labelled as $\Pi_{a_1,a_6}$, where $a_n$ are eigenvalues of the generators $X_n$
\begin{equation}
    \begin{split}
        &\Pi_{-1,0}= -i\Pi_1+\Pi_2 \; , \\ 
        &\Pi_{1,0}= i\Pi_1+\Pi_2  \; ,\\
        &\Pi_{-\oneh,-\oneh}= \frac{-\Pi_3+i\Pi_4+i\Pi_6+\Pi_7}{2}  \; ,\\
        &\Pi_{\oneh,-\oneh}= \frac{\Pi_3+i\Pi_4-i\Pi_6+\Pi_7}{2}  \; ,\\
        &\Pi_{-\oneh,\oneh}= \frac{\Pi_3-i\Pi_4+i\Pi_6+\Pi_7}{2}  \; ,\\
        &\Pi_{\oneh,\oneh}= \frac{-\Pi_3-i\Pi_4-i\Pi_6+\Pi_7}{2}  \; ,\\
        &\Pi_{0,-1} = -\frac{1}{\sqrt{2}} \Pi_5 -i\sqrt{\frac{3}{2}}\Pi_8+\Pi_9  \; , \\
        &\Pi_{0,1} = -\frac{1}{\sqrt{2}} \Pi_5 +i\sqrt{\frac{3}{2}}\Pi_8+\Pi_9  \; , \\
        & \Pi_{0,0}=\sqrt{2}\Pi_5 + \Pi_9  \; .
    \end{split}
\end{equation}
From these expressions it can be shown that the operation of charge conjugation acts on this multiplet as a transformation of $SO(4)$. As a consequence, any $G$-parity is equivalent to an isospin rotation and does not provide selection rules for transition amplitudes. Implications of this feature will be discussed in Sec.~\ref{sec::measurements}.

\section{Perturbative Results}
\label{sec:pt}

In this work we adopt a Wilson-type discretisation of the Dirac action with a clover term, the details of which will be given in Sec.~\ref{sec:lattice_setup}. This choice for the action breaks chiral symmetry. The critical mass, i.e. the value of the bare mass of a fermion corresponding to a vanishing renormalised mass can be first estimated in perturbation theory, suggesting values for the numerical simulations, and providing first insights on the multi-representation dynamics. The perturbative expansion of the Wilson action generates, for each representation, the same vertices that appear in lattice QCD up to group theoretical factors, making it easy to generalise the existing result. To this end, we will refer to the calculation of the critical mass of Wilson fermions at two loops \cite{Follana:2000mn} in lattice QCD, the cactus resummation at one loop \cite{Panagopoulos:1998xf} and their generalisation to a generic representation of $SU(N)$ \cite{DelDebbio:2008wb}. These results will be extended to the case of multiple representations in the reminder of this section. 

\subsection{Multiple representations}
\label{sec:multirep_pt}

In this section we analyse the effect of multiple representations in the computation of the fermionic self-energy. The motivation is to gain insights about the way a representation can affect the other. For simplicity we use, for this task, Wilson-type fermions, delegating the discussion of the clover term and cactus improvement to the next section where we estimate critical masses.

For a given representation, we write the perturbative expansion of the one particle irreducible two-point function as
\begin{equation}
    \Sigma(p,g_0,m_0) = g_0^2 \Sigma^{(1)}(p,m_0) + g_0^4 \Sigma^{(2)}(p,m_0) + \Op(g_0^6) \; .
\end{equation}
The one-loop contribution $\Sigma^1$, for instance, takes contributions from two diagrams, a tadpole and a sunset. It is useful to parametrise $\Sigma^1$ in terms of powers of the lattice spacing,
\begin{equation}
    \Sigma^{(1)}(p) = \frac{\Sigma_\mathrm{a}}{a}+ i \slashed{p} \Sigma_\mathrm{b} \; .
\end{equation}
Imposing the vanishing of the renormalised mass for the fermion yields an expression for the critical mass at one loop
\begin{equation}\label{eq:mc_generic}
    m_\mathrm{c}^{(1)} = \frac{g_0^2 \Sigma_\mathrm{a}}{\mathrm{a}} =  \frac{g_0^2 \Sigma^{\mathrm{(1)}}(p=0)}{a}  \; .
\end{equation}
The same can be done order by order in powers of the coupling $g^2_0$, which is related to the usual lattice coupling by 
\begin{equation}
    \beta = \frac{2N}{g_0^2} \; .
\end{equation}
The result in a given representation $R$ can be written in terms of the contribution of each diagram~\cite{Follana:2000mn}
\begin{equation}\label{eq:one_loop_selfenergy}
    \Sigma^{(1)}(R) = 2C_2(R) \left[ c_1^{(1)} +  c_2^{(1)} \right] \; , \;\;\;\;\;\;\;\; c_1^{(1)} +  c_2^{(1)}  = -0.162857058711(2) \; .
\end{equation}
This result, plugged into Eq.~\eqref{eq:mc_generic}, provides the one-loop estimate of the critical mass. However, it contains no information about the multi-representation dynamics, for which we need at least the $\Op(g_0^4)$ result. In Wilson lattice QCD, $\Sigma^{(2)}$ takes contribution from 26 diagram \cite{Follana:2000mn}.
\begin{figure}[b!]
    \centering
    \includegraphics[width=0.6\textwidth]{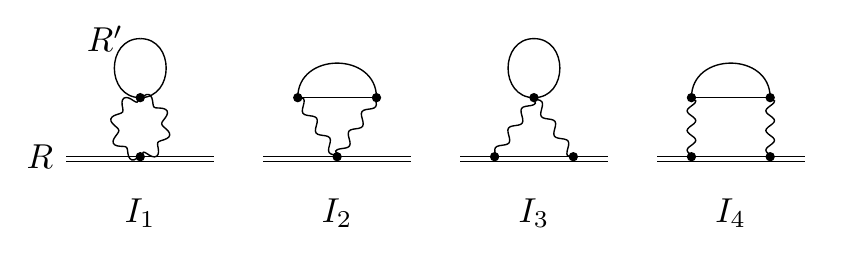}
    \caption{Two-loop diagrams in which the representation $R'$ (solid line) can contribute to the self energy, and therefore the critical mass, of the fermion in the representation $R$ (double solid line). These diagrams, apart from group theoretical invariants, depend on factors that have been computed in Ref.~\cite{Follana:2000mn}.}
    \label{fig:ptdiagrams}
\end{figure}
If the theory has fermions in two different representation $R$ and $R'$, there are four additional diagrams contributing to the self-energy of $R$ due to loops of $R'$. These are shown in Fig.~\ref{fig:ptdiagrams} and their contribution is the same as their single-representation counterparts~\cite{Follana:2000mn} up to some group theory factors. The diagrams $I_1$ and  $I_2$ in Fig.~\ref{fig:ptdiagrams} must be evaluated together in order to be infrared finite. Their value is
\begin{equation}
    I^{R}_1+I^{R}_2 = n_f(R') 4 C_2(R) T_{R'} \, c_1^{(2)},
\end{equation}
where $c^{(2)}_1=0.00079263(8)$ \cite{Panagopoulos:1998xf} is representation independent, and $n_f$ is the number of fermions. Similarly, $I_3+I_4$ gives the infrared-finite result
\begin{equation}
I_3^R + I_4^R = n_f(R') 4 C_2(R) T_{R'} \; c^{(2)}_2 \,  ,
\end{equation}
where $c^{(2)}_2 = 0.000393556(7)$. The two-loop part of the self energy for the representation $R$ due to the presence of $R'$ is then
\begin{equation}\label{eq:multirep_2loop}
     \Sigma_{\mathrm{multi-rep}}^{(2)}(R'\rightarrow R) = n_f(R') 4 C_2(R) T_{R'}\; 0.00118619(9) \;.
\end{equation}
We then add to this contribution the 26 single-representation diagrams $\Sigma^{(2)}_{one-rep}$ \cite{DelDebbio:2008wb}
\begin{equation}\label{eq:singlerep_2loop}
    \Sigma^{(2)}_{\mathrm{one-rep}}(R\rightarrow R) = C_2(R) N \, k_1 + 2C_2(R) T_R n_f(R)\, k_2 + C_2(R) C_2(\mathrm{Fund})\, k_3 + C_2(R)^2 \,k_4 \; ,
\end{equation}
with
\begin{equation}
    k_1 = -0.001940(6) \, , \;\;\;\;\; k_2 = 0.00237236(16) \, , \;\;\;\;\; k_3=-0.081429(8) \; , \;\;\;\;\; k_4 = 0.01516325(12) \; .
\end{equation}
The total two-loop self energy is given by the sum of Eqs.~\eqref{eq:multirep_2loop} and~\eqref{eq:singlerep_2loop}
\begin{equation}
\Sigma^{(2)}(R)  = C_2(R) N \, k_1 + 2C_2(R)\left[ T_R n_f(R) +  T_{R'} n_f(R') \right] \, k_2 + C_2(R) C_2(\mathrm{Fund})\, k_3 + C_2(R)^2 \,k_4 \; .
\end{equation}
Unsurprisingly if $R=R'$ the extra term is equivalent to the addition of extra flavor content. 

We can now list one and two-loop results for $N=4$, $n_f(\mathrm{Fund})=n_f(\mathrm{2AS})=2$
\begin{equation}
\begin{split}
    & \Sigma^{(1)}(\mathrm{Fund}) = -0.610713970166(8) \, , \;\;\;\;\; \Sigma^{(1)}(\mathrm{2AS}) = -0.814285293555(10) \; ,\\
\end{split}
\end{equation}
\begin{equation}
    \Sigma^{(2)}(\mathrm{Fund}) = -0.220826(53) \; , \;\;\;\;\; \Sigma^{(2)}(\mathrm{2AS}) = -0.270743(71) \; ,
\end{equation}
\begin{equation}
    \Sigma_{\mathrm{multi-rep}}^{(2)}(\mathrm{2AS}\rightarrow \mathrm{Fund}) = 0.0213512(14) \; , \;\;\;\;\; \Sigma_{\mathrm{multi-rep}}^{(2)}(\mathrm{Fund}\rightarrow \mathrm{2AS}) = 0.0355854(24) \; .
\end{equation}
The multi-rep contributions alone are small compared to the rest of the terms, providing about $10-13\%$ of the two-loop part. Although these results are only perturbative, they will find a non-perturbative counterpart in Sec.~\ref{sec:chiral_limit}.

\subsection{Critical mass}
\label{sec:critical_mass}

A perturbative prediction that compares to our numerical results can be obtained by including the clover term into the action, resulting into an additional interaction vertex. The prediction can be further improved by resumming an infinite series of a specific type of gauge invariant (\textit{cactus}) diagrams~\cite{Panagopoulos:1998xf}. For this analysis, we consider one-loop results.

The critical mass $m^{(1)}_\mathrm{c}(R)$ of a Wilson fermion can be computed from Eq.~\eqref{eq:one_loop_selfenergy}. Considering the clover term yields new contributions in powers of $c_{\mathrm{sw}}$~\cite{PhysRevD.86.014505},
\begin{equation}\label{eq:clover_crit}
    m_\mathrm{c}(R) = \frac{g_0^2 C_2(R)}{16 \pi^2} \left( \epsilon_0 + \epsilon_1 c_{\mathrm{sw}} + \epsilon_2 c_{\mathrm{sw}}^2 \right) \; ,
\end{equation}
\begin{figure}[t!]
    \centering
    \includegraphics[width=0.5\textwidth]{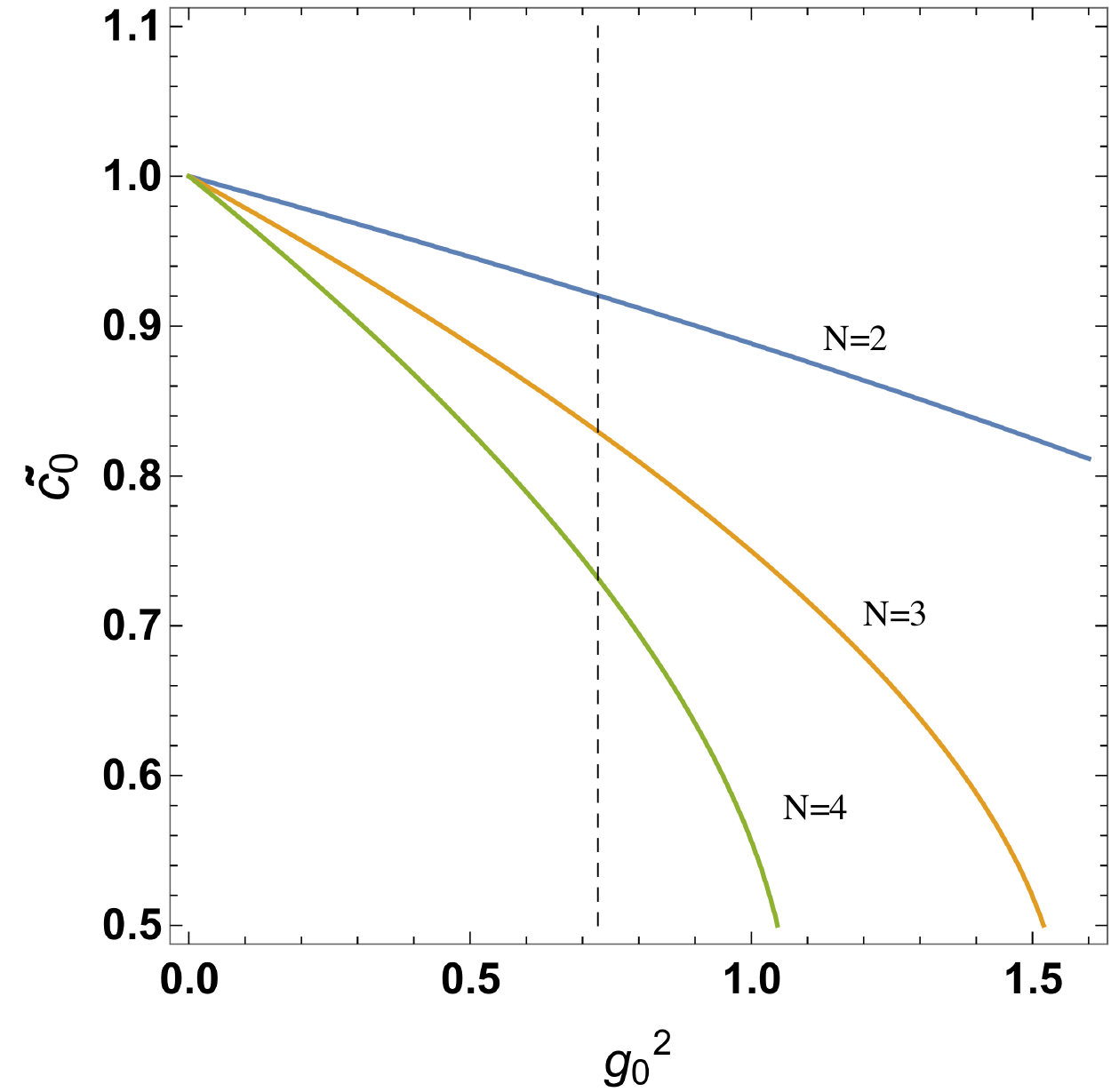}
    \caption{The solution of Eq.~\eqref{eq:defining_ctilde_cactus} for $N=2,3,4$ and $\beta=11$, the value used in the numerical simulations, as will be described in Sec.~\ref{sec:lattice_setup}.}
    \label{fig:cactus}
\end{figure}
where the coefficients $\epsilon_i$ can be read from Table 2 of Ref.~\cite{PhysRevD.86.014505}. Concerning the Wilson term, the cactus resummation is performed by rescaling the self-energy $\Sigma \rightarrow \Sigma/\tilde{c_0}$, where $\tilde{c_0}$ is a function of $N$ and $g_0^2$ but not of the representation. $\tilde{c}_0$ can be found by solving the following equation,
\begin{equation}\label{eq:defining_ctilde_cactus}
    \begin{split}
	& u e^{-u(N-1)/(2N)} \left[ \frac{N-1}{N} L^1_{N-1}(u) + 2L^2_{N-2}(u) \right] = \frac{g_0^2(N^2-1)}{4} \; ,\\
	&  \tilde{c}_0 \equiv \frac{g_0^2}{4u} \; ,
	\end{split}
\end{equation}
where $L_N^\alpha$ are generalised Laguerre polynomials of degree $N$. The solution for $\beta=11$ and $N=2,3,4$ is shown in Fig. \ref{fig:cactus} as the intersection between the various curves and the vertical line. For $SU(4)$ we find
\begin{equation}
 \tilde{c}_0 = 0.731607 \;.
\end{equation}
The resummation with the clover improvement term corresponds to also rescale $g_0^2\rightarrow g_0^2/\tilde{c}_0$ and $c_{\mathrm{sw}} \rightarrow c_{\mathrm{sw}} \tilde{c}_0$ \cite{PhysRevD.74.074503}. The prediction for the critical mass is then
\begin{equation}
    m_{\mathrm{c}}^{\mathrm{1-loop+cactus}}(R) = \frac{g_0^2}{\tilde{c}_0} \frac{C_2(R)}{16 \pi^2} \left( \epsilon_0 + \epsilon_1 c_{\mathrm{sw}} \tilde{c}_0 + \epsilon_2 c_{\mathrm{sw}}^2 \tilde{c}_0^2 \right) \; .
\end{equation}
The numerical values for $SU(4)$ with two fundamental and two antisymmetric fermions are listed in Table \ref{tab:critical_masses}. We find the one-loop result with cactus resummation to be the closest one to non-perturbative results.
 \begin{table}[h!]
     \centering
     \begin{tabular}{c | c c c c c}
          $\;$ &  W.  & W. (2 loop) & W.c.  & W.+cactus & W.c.+cactus \\[-1em]\\ \hline
          \\[-0.8em]
          $am_c^{(\mathrm{F})}$  & $-0.4442$ & $-0.5609$ & $-0.2762$ & $-0.6063$ &  $-0.4524$   \\
          \\[-0.8em]
          $am_c^{(\mathrm{2AS})}$  &  $-0.5922$ & $-0.8027$ & $-0.3683$  & $-0.8089$ &  $-0.6032$   \\[-1em]\\ \hline
     \end{tabular}
     \caption{Critical masses for different lattice actions and improvements. W. is an abbreviation for Wilson fermion, W.c. for Wilson clover. When not specified, the results are obtained at 1 loop. These values are to be compared with the non-perturbative results of Sec.~\ref{sec:chiral_limit}.}
     \label{tab:critical_masses}
 \end{table}

\section{Lattice Setup}
\label{sec:lattice_setup}

\begin{figure}[t]
    \centering
    \includegraphics[width=0.48\textwidth]{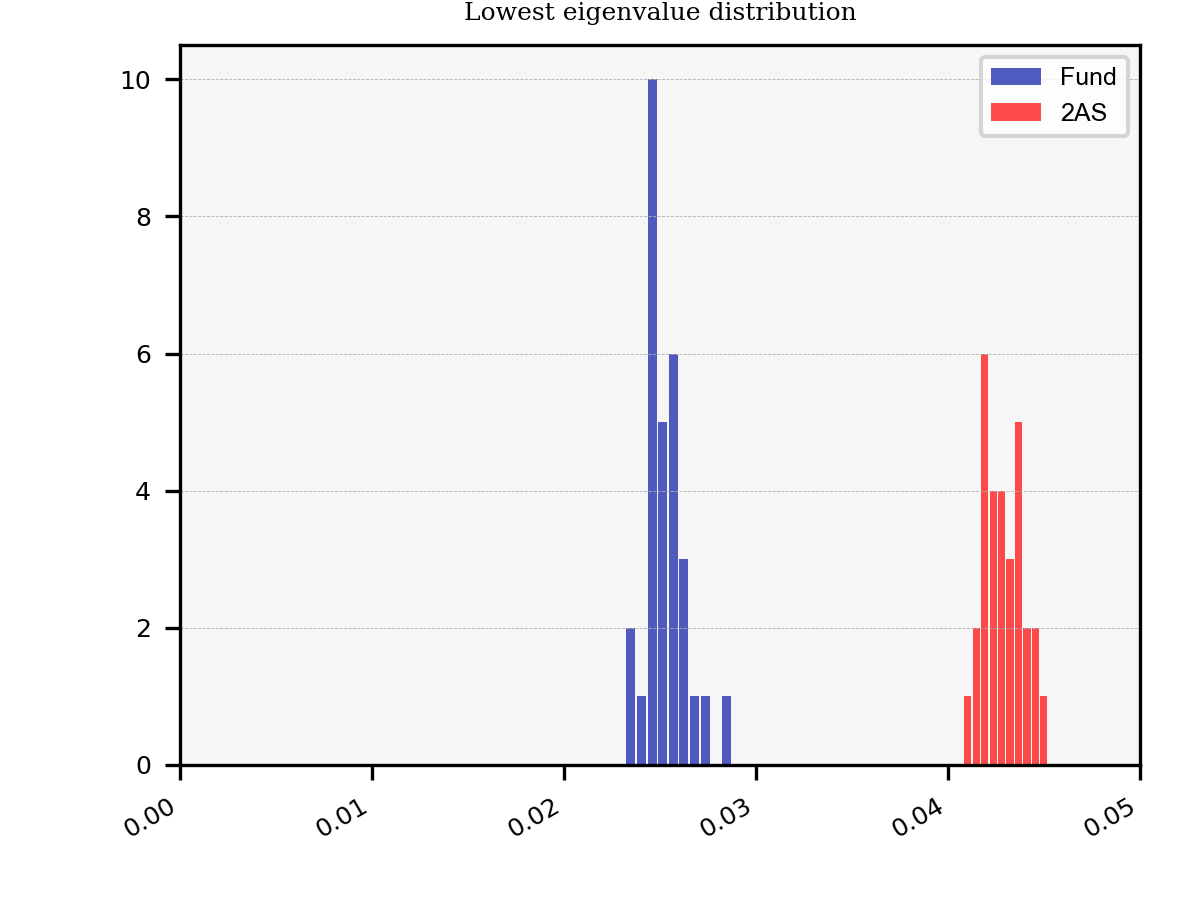}
    \includegraphics[width=0.48\textwidth]{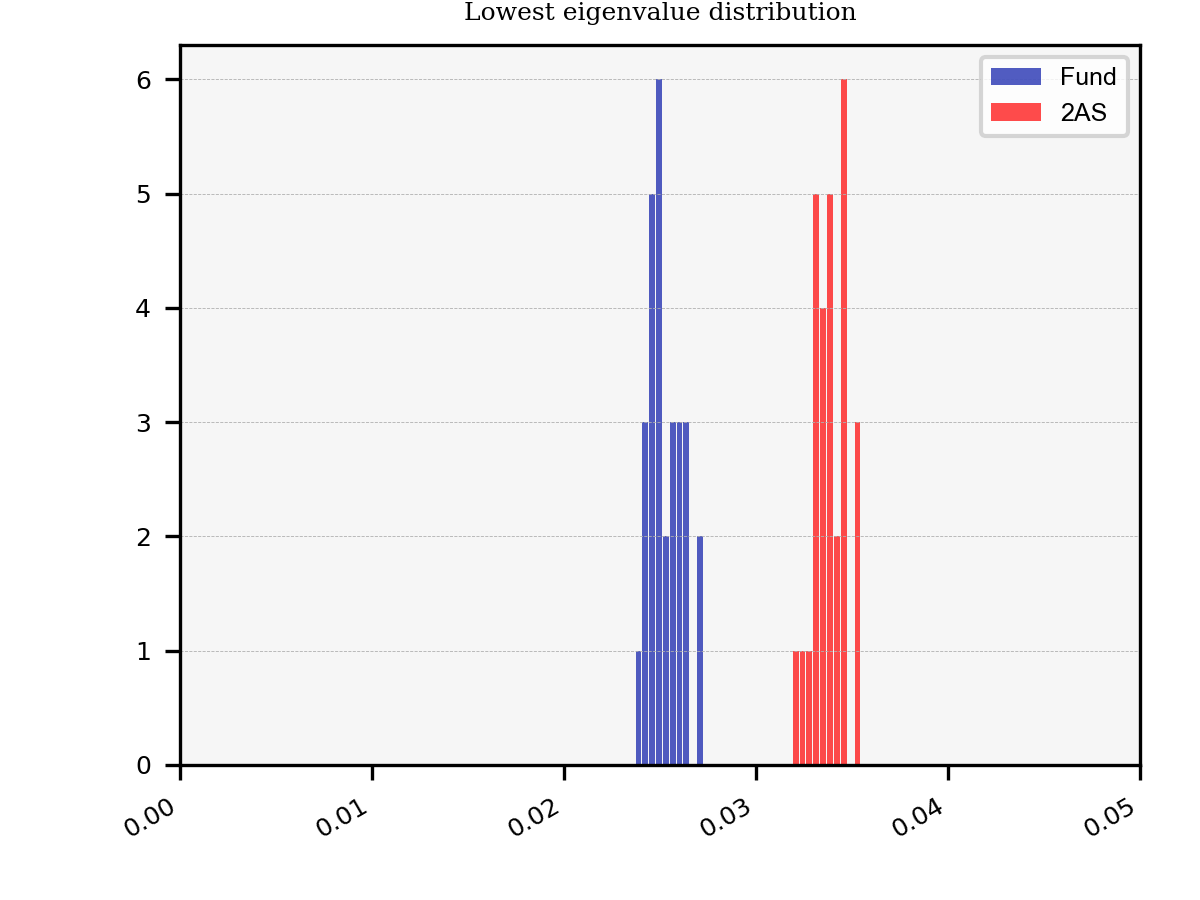}
    \caption{Distribution of the lowest eigenvalue of the Wilson-clover operator for bare parameters $am_0^{\mathrm{F}}=-0.45$, $\beta=11$ and different masses for the 2AS fermions: $am_0^{\mathrm{2AS}}=-0.59$ on the left and $am_0^{\mathrm{2AS}}=-0.60$ on the right. Both distributions are far enough from the origin to ensure no exceptional configurations enter the ensemble.}
    \label{fig:m56-59_lowest_eig_distrib}
\end{figure}

The lattice setup has already been discussed in Ref.~\cite{Cossu_2019}, and here we only recall the main ideas. We generate gauge configurations by using the HMC algorithm~\cite{Duane:1987de} with a multi-level second order integrator~\cite{OMELYAN2003272}. The lattice action can be decomposed into gauge and fermionic part. The latter is further decomposed into contributions from each representation~\cite{Del_Debbio_2010}
\begin{equation}
    S = S_\mathrm{g} + S_\mathrm{f}\; , \;\;\;\;\;\; S_\mathrm{f} = S^{({\mathrm{Fund}})} + S^{({\mathrm{2AS}})} \; .
\end{equation}
For the gauge part we use the Wilson plaquette action
\begin{equation}
    S_\mathrm{g} = \frac{\beta}{N_c} \sum_{x} \sum_{\mu < \nu} \; \text{Re}\; \text{Tr} \left\{ 1 - \mathcal{P}_{\mu\nu}(x) \right\} \; .
\end{equation}
The fermionic action for the representation $R$ is
\begin{equation}
    S_\mathrm{f}^{R} = \sum_x \bar{\psi}^{R}(x) \, D^R_{x,y} \, \psi(x)^{R} \; .
\end{equation}
We adopt a Wilson discretisation for the Dirac action $D^R$ in each representation $R$, plus a clover improvement term
\begin{equation}
    D^R = D_{\mathrm{Wilson}}^R + 
    D_{\mathrm{clover}}^R \; .
\end{equation}
The Wilson term in position space is
\begin{equation}
    D^R_{x,y} = \delta_{x,y} - \kappa^R \sum_{\mu=1}^4 \left[ \left( \idd - \gamma_\mu \right) U^R_\mu(x) \delta_{x+a\hat{\mu},y} + \left( \idd + \gamma_\mu \right) U_\mu^R\,\da(y) \delta_{x-a\hat{\mu},y} \right] \; ,
\end{equation}
where
\begin{equation}
    U^R_\mu(x) = \exp(i\,\omega^a_\mu(x) \, T_a^R) \; ,
\end{equation}
and $\kappa^R$ is related to the bare mass $m_0^R$ of the fermion in the representation $R$
\begin{equation}
    \kappa^R = \frac{1}{2(am_0^R+4)} \; .
\end{equation}
The order $a$ clover improvement term is
\begin{equation}
    \left( D^R_{\mathrm{clover}} \right)_{x,y} = \frac{ia}{2} \, c^R_{\mathrm{sw}}(g_0^2) \, \kappa^R \, \sum_{\mu\nu} \tilde{F}^R_{\mu\nu}(x) \, \sigma_{\mu\nu} \delta_{x,y} \; ,
\end{equation}
where $\sigma_{\mu\nu}=\frac{i}{2}[\gamma_\mu , \gamma_\nu]$ and 
\begin{equation}
    \tilde{F}_{\mu\nu}(x) = \frac{1}{8} \left[ Q_{\mu\nu}(x) - Q_{\nu\mu}(x) \right] \; , \;\;\;\;\; Q_{\mu\nu}(x) = Q_{\nu\mu}\da(x) \; ,
\end{equation}
$Q_{\mu\nu}(x)$ being the clover combination of plaquettes around the point $x$~\cite{SHEIKHOLESLAMI1985572}. The improvement coefficient $c^R_{\mathrm{sw}}$ can be computed in powers of the coupling,
\begin{equation}
    c^R_{\mathrm{sw}}(g_0^2) = 1 + c_{\mathrm{sw}}^{R\,(1)} \, g_0^2 + \Op(g_0^4) \; .
\end{equation}
A discussion on the $O(g_0^2)$ coefficient can be found in~\cite{Musberg_2013}. In this work, we set the coefficient to its tree level value, $c^R_{\mathrm{sw}}=1$, for both representations. This choice for the discretisation action breaks explicitly chiral symmetry, resulting in an additive term to the renormalisation of the fermion masses. Simulations at the chiral point are performed indirectly by extrapolating the value of the critical masses $m_\mathrm{c}^R$, i.e. the value of the bare masses at which each fermion has a renormalised vanishing mass
\begin{equation}\label{eq:critical_k}
    am^R = am_0^R - am_\mathrm{c}^R = \oneh \, \left( \frac{1}{\kappa^R} - \frac{1}{\kappa^R_\mathrm{c}} \right) \; .
\end{equation}
Direct simulations at the chiral point are in fact impossible due to the spectral properties of the Wilson operator. The lowest eigenvalue of the Wilson operator approaches zero in the chiral limit, meaning its inverse becomes increasingly ill-conditioned. When extrapolating to the critical point (see Eq.~\eqref{eq:critical_k}) one needs to care that exceptional configurations do not occur in the gauge average. These are configurations especially close to the chiral point that can jeopardise the inversion of the fermionic operator. For this reason, we monitor the gauge distribution of the lowest eigenvalues of the Wilson-clover operator~\cite{DelDebbio:2005qa} as in Fig.~\ref{fig:m56-59_lowest_eig_distrib}, making sure that they remain sufficiently far from the origin. In order to parametrise the breaking of chiral symmetry, we compute the PCAC mass, which yields a definition of the quark mass through the axial Ward identity.

Our starting point is the work done in \cite{Cossu_2019}, where the space of bare parameters for this model has been explored. The lightest ensemble of that work is present in this analysis under the name A0. Details about the ensembles generated in this work are found in Appendix \ref{app:ensembles}. We have performed simulations for increasingly lighter masses, allowing the extrapolation of chiral points for both the fermionic representations. We generate gauge configurations at single lattice spacing, using $\beta=11$, and at a single volume corresponding to a lattice of dimensions $L^3 \times T = 16^3 \times 32$.
For the production of gauge configurations and the measurements of observables, we use the software Grid~\cite{Boyle:2015tjk} and Hadrons~\cite{antonin_portelli_2022_6382460}.

\section{Observables}
\label{sec:observables}

In this section, we give a description of the observables that are targeted in this work. These include two point functions that will enable us to compute parameters for the chiral symmetry breaking, mesonic masses and smeared spectral densities. We compute correlation functions of the following interpolators which are defined for each representation $R$
\begin{equation}\label{eq:interpolators_generic}
\begin{split}
    &   O_{\;\mathrm{P}}^{R}(x) = \bar{\psi}^R(x)\,  \gamma_5 \,\psi^R(x)\\
    &   O_{\;\mathrm{A}}^{R}(x) = \bar{\psi}^R(x) \,\gamma_\mu \gamma_5 \, \psi^R(x)\\
    &   O_{\;\mathrm{V}}^{R}(x) = \bar{\psi}^R(x)\,  \gamma_i \, \psi^R(x) \, , \;\;\;\;\; i=1,2,3
    \end{split}
\end{equation}
In this work we only consider isospin-vector operators, which are used to build two-point functions
\begin{equation}\label{eq:generic_lattice_correlator_2pt}
    C_{ab}^{R}(t) = \frac{1}{L^3} \sum_{\vet{x}} \braket{O^{R}_a(\vet{x},t) \bar{O}^{R}_b(0)} \; , \;\;\;\;\; a,b = \mathrm{P},\mathrm{A},\mathrm{V} \; .
\end{equation}
The two-point functions encode information about finite volume matrix elements and energy levels. This can be seen by expanding the previous expression on a complete set of states:\footnote{In a finite volume $L^3\times T$, the correlators depend both on $L$ and $T$. Since the dependence on $T$ is exponentially suppressed~\cite{Bulava:2021fre}, we will omit to label its dependence. The spatial volume $L$, on the other hand, drastically modifies the nature of certain observables and its dependence will be shown explicitly in this section.}
\begin{equation}\label{eq:c_expanded_cosh}
    C_{ab}^{R}(t) = \sum_n \left( e^{-tE_n(L)} + e^{(-T+t)E_n(L)} \right) \, \frac{\braket{0|O^{R}_a(0)|n}_L \braket{n|\bar{O}^{R}_b(0)|0}_L}{2E_n(L)} \; .
\end{equation}
While we will assume, to slim the notation, a positive sign between the exponentials in the previous expression, it is understood that this may vary depending on the quantum numbers $a,b$. The energy of the ground state with given quantum numbers can be obtained by the asymptotic behaviour in the Euclidean time of the corresponding correlator. For example, the effective mass of a pseudoscalar meson can be computed from the pseudoscalar-pseudoscalar correlator
\begin{equation}\label{eq:cosh_mass}
    aM^{R}_{\mathrm{PP}}(t) = \cosh^{-1}\left[ \frac{ C^{R}_{\mathrm{PP}}(t+a) + C^{R}_{\mathrm{PP}}(t-a)}{2C^{R}_{\mathrm{PP}}(t)} \right] \; .
\end{equation}
Similarly the PCAC mass, defined through the axial Ward identity, is obtained from the pseudoscalar and axial correlators
\begin{equation}
    am^{R}_{\mathrm{PCAC}} = \frac{\oneh (\de_t+\de^*_t) C^{R}_{\mathrm{AP}}(t)}{2 C^{R}_{\mathrm{PP}}} \; ,
\end{equation}
which has $\Op(a)$ effects for our choice of the unimproved pseudo-axial operator. 

\begin{figure}[t]
    \centering
    \includegraphics[width=0.32\textwidth]{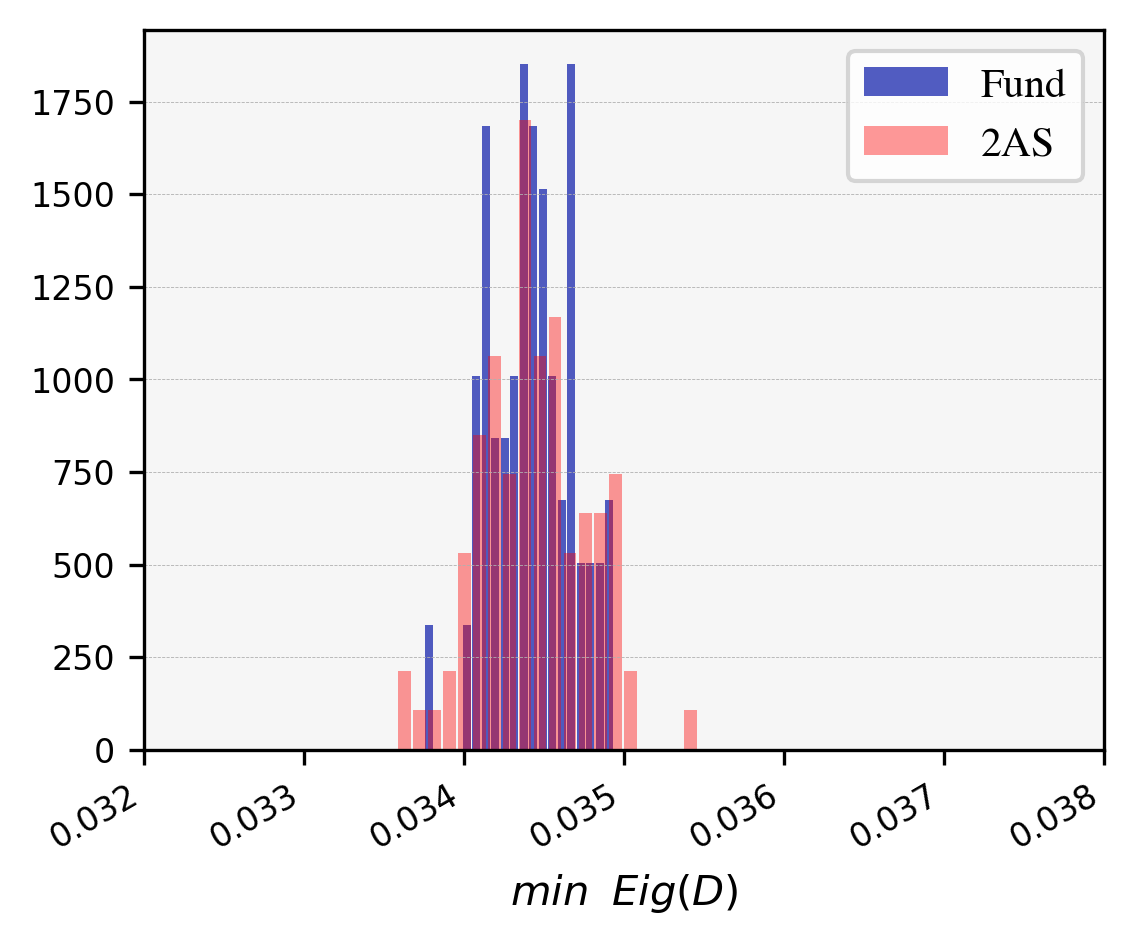}
    \includegraphics[width=0.32\textwidth]{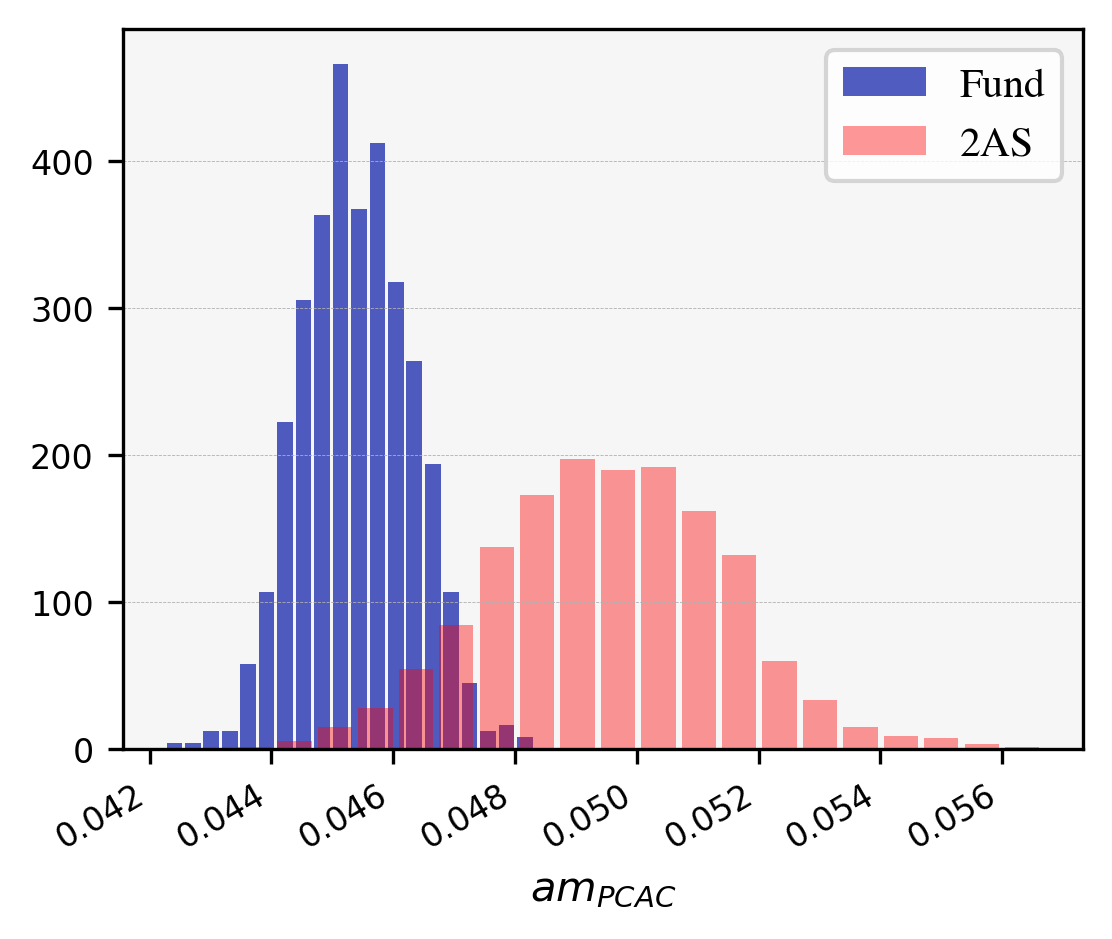}
    \includegraphics[width=0.32\textwidth]{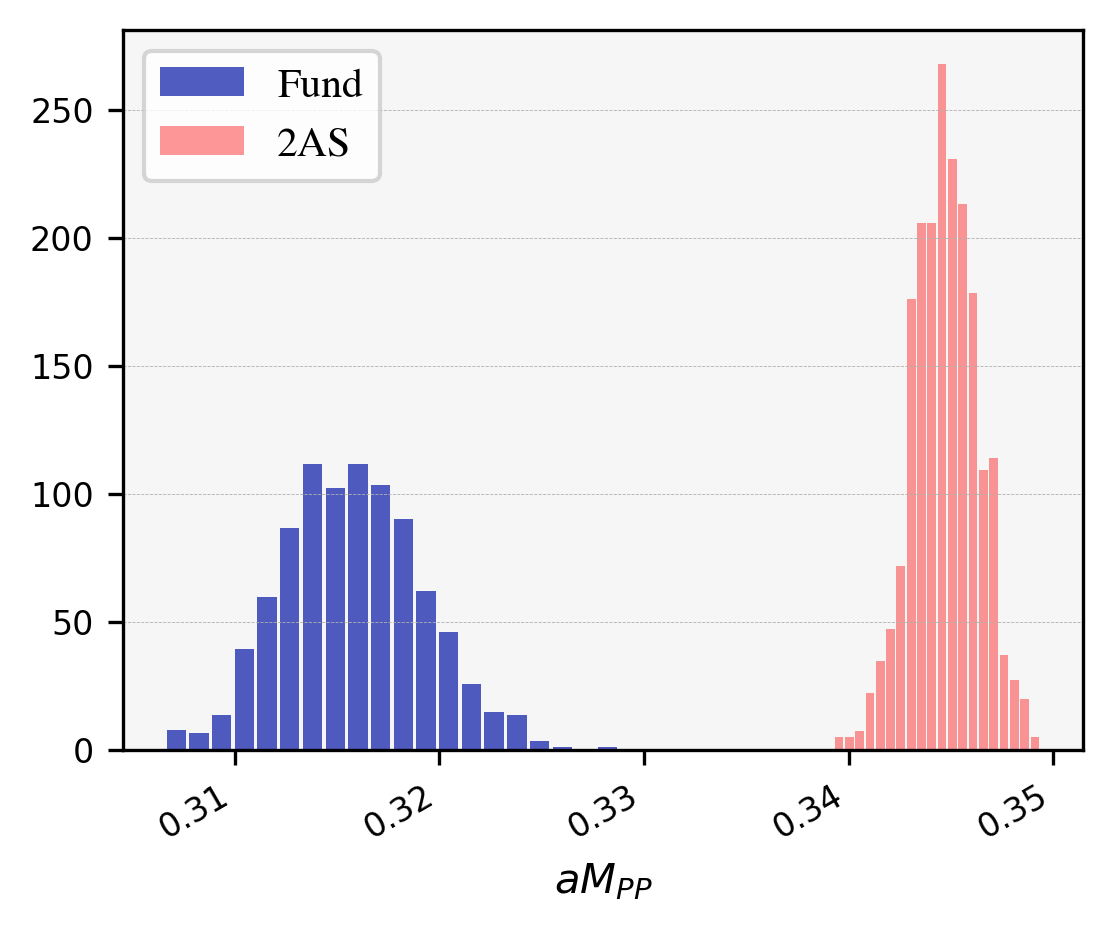}
    \caption{Probability densities for the lowest eigenvalue of the fermionic operator (left), the PCAC mass (centre) and the pseudoscalar mass (right) for both representations. These values are especially interesting since they are based on the ensemble S0, where we find compatible values for the lowest eigenvalues and, within $1\sigma$, the PCAC masses. Nonetheless, a difference arises in the pseudoscalar masses. The distributions of the PCAC and the pseudoscalar masses are obtained from a resampled set of configurations.}
    \label{fig:symmetric_ensemble}
\end{figure}

In the chiral limit, the pseudoscalar mass $aM_{\mathrm{PP}}^{R}$ for each representation vanishes. In our ensembles we generated configurations describing mesons of antisymmetric fermions being generally heavier: this can be understood from Fig. \ref{fig:symmetric_ensemble} where we show the gauge distributions of the lowest eigenvalue of the fermionic operator, the fermionic masses and the pseudoscalar masses for both representations from the ensemble $S0$. In this ensemble, the lowest eigenvalues of the Wilson operator are the same, and the fermionic masses are also compatible within one $\sigma$. Nonetheless, an unambiguous gap appears in the masses of the mesons. The Gell-Mann--Oakes--Renner relation predicts this behaviour to be the result of different chiral condensates and pseudoscalar decay constants for the two representations.  

Ground state energies can be estimated from lattice correlators according to Eq.~\eqref{eq:cosh_mass}. Other energy levels and matrix elements can be estimated by fitting sums of exponentials, as in Eq.~\eqref{eq:c_expanded_cosh}. The extraction of excited states from these fits is in general hindered by the increasing number of degrees of freedom that are needed in order to perform the fit when more excited states are being targeted, with a given number of data points. This becomes particularly problematic when dealing with highly correlated data, which limits the effective number of degrees of freedom. Moreover, as the infinite volume limit is approached, the spectrum above the multi-particle threshold becomes continuous and resolving energy levels becomes exponentially harder. For these reasons, it is desirable to have many correlators in order to perform simultaneous fits. Alternatively, variational methods such as the generalised eigenvalue problem (GEVP) are a well established way to obtain finite volume spectra~\cite{collaboration_2009}. 

Other observables that allow the extraction of finite volume quantities are spectral densities. These are related to lattice correlators by a Laplace transform
\begin{equation}
    C^{R}_{ab}(t) = \int_0^{\infty} dE \, \rho^{R}_{ab}(E) \,  \left( e^{-tE}+ e^{(-T+t)E} \right) \; .
\end{equation}
Spectral densities contain the same information as lattice correlators, with the difference that for the spectral densities the information is encoded in a function of the energy rather than Euclidean time. A finite volume spectral density $\rho^L_{ab}(E)$ can in fact be expanded as
\begin{equation}\label{eq:specdens_expanded_cosh}
    \rho_{ab}^{L,R}(E) = \sum_n \frac{\braket{0|O^{R}_a(0)|n}_L \braket{n|\bar{O}^{R}_b(0)|0}_L}{2E_n(L)}\, \delta\left( E-E_n(L) \right) \; .
\end{equation}
The distributional nature of the spectral density occurs due to the finite volume of the simulation and it is mostly lost in the continuum limit, where above the multi-particle threshold the spectral density becomes continuous, while it retains Dirac deltas in presence of single-particle or bound states. In order to treat this distributional character we smear the finite volume spectral density \cite{Hansen:2017mnd} with a Gaussian function, $\Delta_\sigma(E) = \exp{(-E^2/2\sigma^2)}/\sqrt{2\pi}\sigma$
\begin{equation}\label{eq:gau_sm}
    \rho^{L,R}_{\sigma , ab}(E) = \int_0^\infty dE\p \, \Delta_\sigma(E-E') \, \rho_{ab}^{L,R}(E') \; ,
\end{equation}
so that the smeared spectral density $\rho^{L,\,R}_{\sigma , ab}(E)$ is a continuous function even at finite $L$. 

In this work, we will focus on the extraction of the finite volume energies and matrix elements from smeared spectral densities. This task does not require us to take the infinite volume limit, neither to remove the regularisation by extrapolating at zero smearing radius $\sigma$. 
By rewriting Eq.~\eqref{eq:specdens_expanded_cosh} for the smeared spectral density,
\begin{equation}\label{eq:smeared_spectraldens_explicit}
    \rho_{\sigma , ab}^{L,R}(E) = \sum_n \frac{\braket{0|O^{R}_a(0)|n}_L \braket{n|\bar{O}^{R}_b(0)|0}_L}{2E_n(L)}\, \Delta_\sigma \left( E-E_n(L) \right) \; ,
\end{equation}
it is clear that this function can be fitted against sums of Gaussians in order to obtain the energies and the overlaps, similarly to how Eq.~\eqref{eq:c_expanded_cosh} is commonly used to extract the same quantities by fitting sums of exponentials. This offers a dual picture with quite different features that will be described in the next sections.

\section{Chiral Limit}
\label{sec:chiral_limit}

At the chiral point, both pseudoscalar mesons become massless. The discretisation of the lattice action that we adopt breaks chiral symmetry explicitly. We therefore compute the PCAC masses from the axial Ward identity in order to quantify the breaking of chiral symmetry, and we scan the space of bare parameters until we are able to locate the bare masses for which the PCAC masses vanish. At the chiral point we also expect the Gell-Man--Oakes--Renner relation to be valid, with the pseudoscalar mass squared $(M^{R}_{\mathrm{PP}})^2$ scaling as $m_{\mathrm{PCAC}}^{R}$ for the representation $R$. The correlation functions of fermions in the fundamental representation that we use to extract the masses are more affected by autocorrelation, which translates into usually larger statistical errors for the masses of the mesons related to that representation.
\begin{figure}[t]
    \centering
    \includegraphics[width=0.495\textwidth]{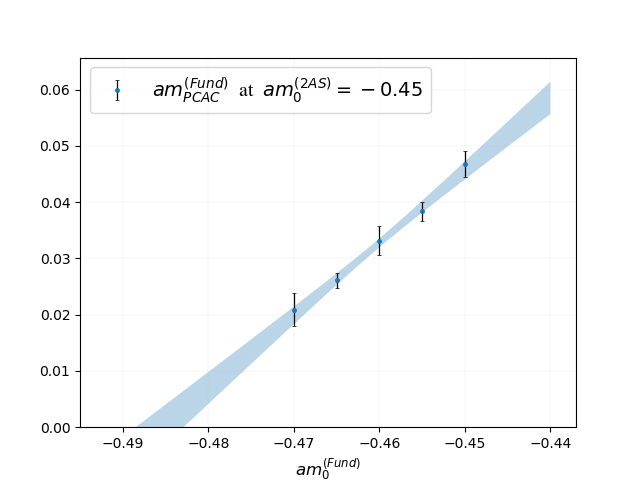}
    \includegraphics[width=0.495\textwidth]{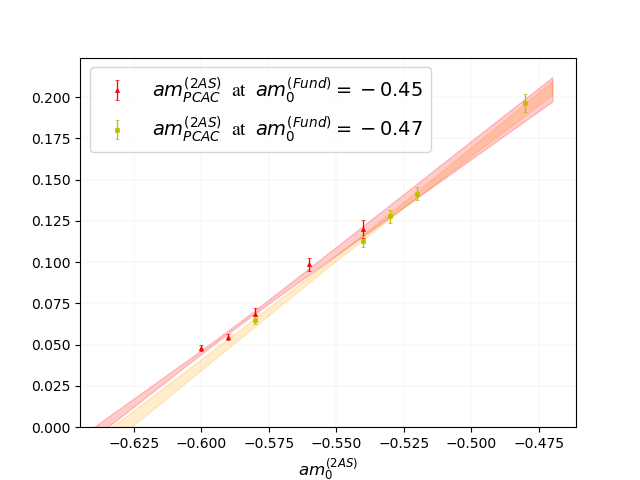}
    \caption{Three chiral extrapolations, one on the left for fundamental fermions and two on the right for antisymmetric fermions. The bands correspond to linear fits. The heavier point in the plot on the left was also present in \cite{Cossu_2019}.}
    \label{fig:chiral_limits}
\end{figure}
The measurements of the PCAC masses are listed in 
Tables~\ref{tab:pcac_fund} and~\ref{tab:pcac_2as} of 
Appendix~\ref{app:sec:masses}. In order to perform a linear extrapolation for the vanishing of the PCAC mass in a given representation, the bare mass of the other must be kept fixed. 
Fig.~\ref{fig:chiral_limits} shows three of these extrapolations. In the left panel, we extrapolate the chiral point for the fundamental fermions by fixing $am_0^{\mathrm{(2AS)}}=-0.45$. In the right panel, the two lines represent different extrapolations for the 2AS chiral point taken at different values of the bare fundamental mass, $am_0^{\mathrm{(F)}}=-0.45$ and $am_0^{\mathrm{(F)}}=-0.47$. The chiral point of antisymmetric fermions does not show a strong response to the shift in the bare mass of the fundamental fermions. This is in line with the perturbative prediction at two loops of Sec.~\ref{sec:pt}, where we have found the critical mass to only mildly depend on the other representation. The chiral points for $\beta=11.0$ are
\begin{equation}
\begin{split}
    & am_\mathrm{c}^{\mathrm{F}}\at_{am_0^\mathrm{(2AS)}=-0.45} = -0.486(3) \ , \\
    & am_\mathrm{c}^{2AS}\at_{am_0^\mathrm{(F)}=-0.45} = -0.637(3) \; , \;\;\;\;\; am_\mathrm{c}^{\mathrm{(2AS)}}\at_{am_0^\mathrm{(F)}=-0.47} = -0.630(4) \; .
    \end{split}
\end{equation}
\begin{figure}[t]
    \centering
    \includegraphics[width=0.49\textwidth]{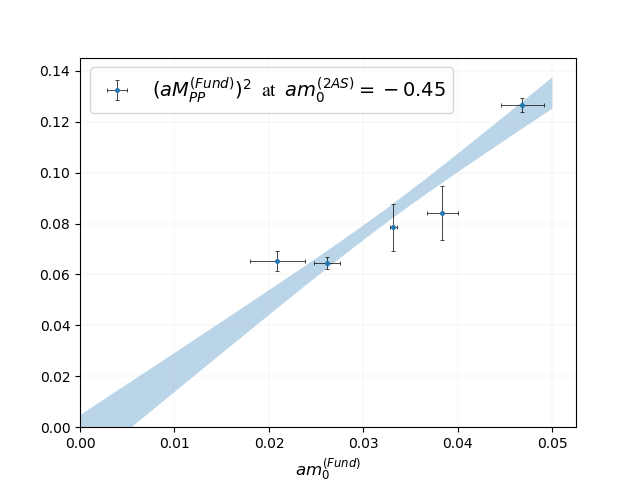}
    \includegraphics[width=0.49\textwidth]{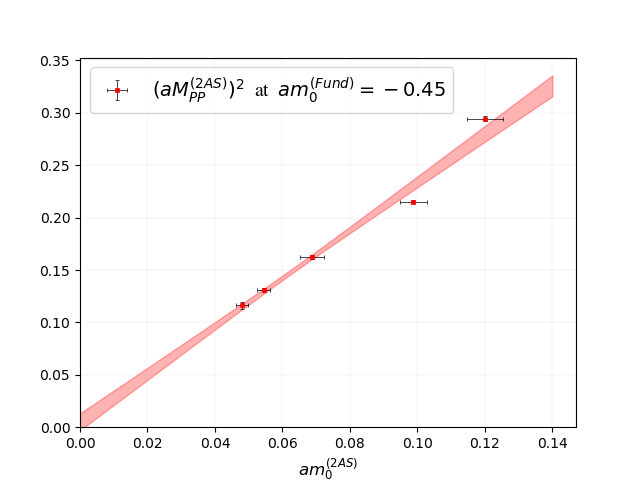}
    \caption{Scaling predicted by the Gell-Man--Oakes--Renner relation, with the squared mass of the pseudoscalar Goldstone boson scaling linearly with the quark mass, here estimated from the PCAC relation. The extrapolation is compatible with the Goldstone bosons becoming massless at the chiral point.}
    \label{fig:GMORs}
\end{figure}
Comparing these values to the perturbative results in Table \ref{tab:critical_masses}, we see that the prediction improved by resumming cactus diagrams is the closest to the non-perturbative values. We can also observe from Tables \ref{tab:pcac_fund} and \ref{tab:pcac_2as} that taking a fermion in the representation $R$ towards the chiral point pushes the fermion in the representation $R'$ to be also lighter. Figure \ref{fig:GMORs} shows the dependence of the pseudoscalar masses $M_{\mathrm{PP}}^{R}$ with respect to the quark masses. For both representations $R$ the scaling is compatible with the one predicted by the chiral Lagrangians, $(M^{R}_{\mathrm{PP}})^2 \sim m^{R}_{\mathrm{PCAC}}$, with the pseudoscalar mesons becoming massless in the chiral limit. The values of the masses are reported in Table \ref{tab:PSmesons}.
\FloatBarrier

\section{Smeared Spectral Densities from Lattice 
Correlators}
\label{sec:backus-gilbert}

The non-perturbative calculation of spectral densities has been receiving increasing attention \cite{Hansen_2019, Hansen:2017mnd, Bulava:2019kbi,  Bulava:2021fre, Gambino:2022dvu, Bruno:2020kyl, Bailas:2020qmv}.
In this work we will analyse spectral densities obtained from lattice correlators, for the first time to our knowledge, in the context of BSM and multi-representation theories. In order to facilitate the discussion, it is useful to recall the computational details of the calculation, aiming to a self-contained discussion.
\subsection{The numerical procedure}\label{sec:backus-gilbert_1}
\begin{figure}[htb]
    \centering
    \includegraphics[width=0.69\textwidth]{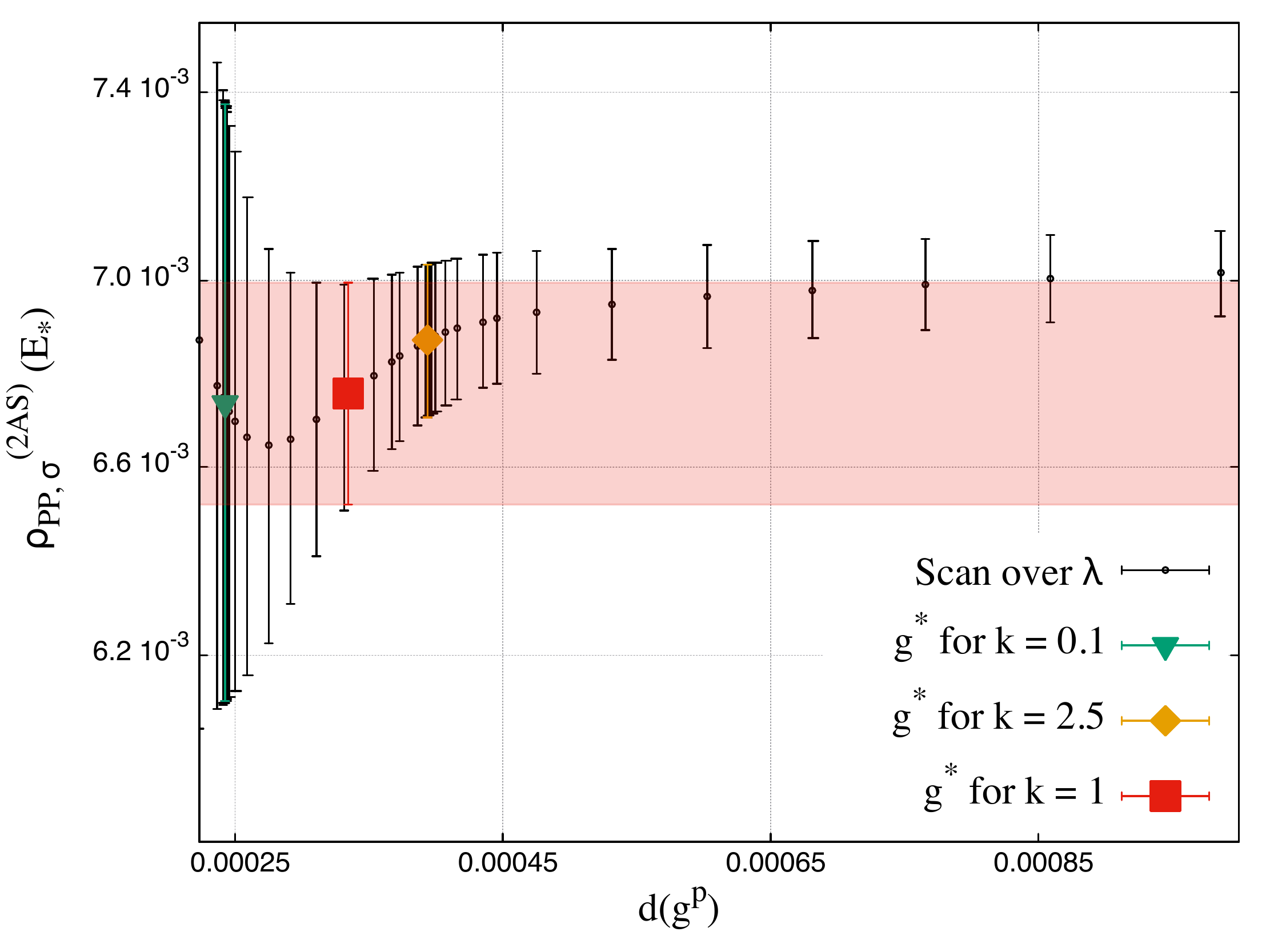}
    \caption{Example of region of algorithmical stability at a given energy $E_\star$. Different values of $\lambda$, which translate into different values of $d(\vet{g}^{\vet{p}})$ on the x-axis, produce predictions for the smeared spectral density $\rho_{ab, \, \sigma}^{R}(E_\star)$ that are compatible within statistical error (black bars). In this case, $R=\mathrm{2AS}$, $\sigma=0.21/a$ and $E_\star \simeq M_{\mathrm{PP}}^{\mathrm{2AS}}$. The green and orange points correspond, according to Eq.~\eqref{eq:ra0}, to values of $k=0.1$ and $k=2.5$ respectively. The red point, extended in the horizontal band, is obtained at $k=1$ and it corresponds to the value of $\lambda$ at which one achieves the optimal balance $ A_\alpha[\vet{g}^{\star}] / A_\alpha[\vet{0}] =\;  B[\vet{g}^{\star}]$. }
    \label{fig:lscan}
\end{figure}

Since we are interested in finite volume quantities we will omit, in this section, the dependence on the spatial volume $L$ that we will assume to be finite. Computing spectral densities $\rho_{ab}^{R}(E)$ from Euclidean correlators $C_{ab}^{R}(t)$ involves an inverse Laplace transform,
\begin{equation}
\label{eq:laplace_transform}
    C_{ab}^{R}(t) = \int_{E_{\mathrm{min}}}^\infty \, dE\, \left( e^{-tE}+e^{(-T+t)E} \right) \, \rho_{ab}^{R}(E) \; ,
\end{equation}
where $E_{\mathrm{min}}$ can range between zero and the energy of the ground state, since $\rho_{ab}^{R}$ vanishes in that interval. The inversion of Eq.~\eqref{eq:laplace_transform} is numerically an ill-posed problem which needs to be regularised. Recalling from Sec.~\ref{sec:observables} that we are interested in the smeared version of $\rho_{ab}^{R}(E)$, it is especially convenient to approach the inverse problem using the Backus--Gilbert type 
regularisation~\cite{10.1111/j.1365-246X.1968.tb00216.x} introduced in~\cite{Hansen_2019}, which yields spectral densities smeared with a chosen smearing function $f(E)$,
\begin{equation}
    \rho_{ab}^{R} [f] = \int_0^\infty \, dE \, f(E) \, \rho_{ab}^{R}(E) \; .
\end{equation}
An important observation is that a fixed smearing kernel is crucial in order to perform fits and extrapolations of the results. The idea of the algorithm is to generate the target smearing kernel, the Gaussian $\Delta_\sigma(E)$ of Eq.~\eqref{eq:gau_sm}, as a linear combination of the same exponentials appearing in the Laplace transform of Eq.~\eqref{eq:laplace_transform},
\begin{equation}\label{eq:kernal_bar}
     \Delta_\sigma(E-E\p) = \sum_{\tau=1}^{\infty} \, g_\tau \, \left( e^{-\tau a E}+e^{(-T+\tau)aE} \right) \; ,
\end{equation}
where $t=\tau a$, $a$ being the lattice spacing. Once the coefficients $g_\tau \equiv g_\tau(\sigma,E\p)$ are known one can simply obtain an estimator for the smeared spectral density, 
\begin{equation}\label{eq:infsum}
    \rho_{ab, \, \sigma}^{R}(E\p) = \sum_{\tau=1}^{\infty} \, g_\tau  \, C_{ab}^{R}(a \tau) \; .
\end{equation}
On the lattice the correlators $C_{ab}^{R}$ are available for a finite number of times, therefore the sum in Eqs.~\eqref{eq:kernal_bar} and~\eqref{eq:infsum} must be truncated at the appropriate cutoff $\tau_{\mathrm{max}}$. Since our lattices have temporal length $T$ and periodic boundaries, we have $a\tau_{\mathrm{max}}=T/2$. The reconstructed smearing kernel,
\begin{equation}
    f(E,\vet{g}) = \sum_{\tau=1}^{\tau_{\mathrm{max}}} g_\tau \,  \left( e^{-\tau a E}+e^{(-T+\tau)aE} \right) \; ,
\end{equation}
will necessarily differ from the Gaussian $\Delta_\sigma(E)$ at finite $\tau_{\mathrm{max}}$, inducing a systematic error on the final result. The computation of the coefficients $\vet{g}$ is achieved through the minimisation of the functional $W_\alpha[\vet{g}]$
\begin{equation}\label{eq:W_def}
    W_\alpha[\vet{g}]  = \frac{A_\alpha[\vet{g}]}{A_\alpha[0]} + \lambda \,B[\vet{g}] \; ,
\end{equation}
where $\lambda$ is a trade-off input parameter that we will discuss later in this section. The functional $A_\alpha[\vet{g}]$, introduced in \cite{Hansen_2019}, measures the difference between the exact smearing kernel and the one we can reconstruct with the available data
\begin{equation}\label{eq:A}
    A_\alpha[\vet{g}] = \int_{E_{\mathrm{min}}}^{\infty} dE \,e^{\alpha aE}  \left| f(E,\vet{g}) -  \Delta_\sigma(E-E\p)  \right|^2 \; .
\end{equation}
The parameter $\alpha<2$ enables the selection between a class of norms in order to measure the distance between the target and the exact function. Choosing larger values of $\alpha$ allows for the integrand to decay faster at high energies. The functional $B[\vet{g}]$ is needed to regularise the problem \cite{10.1111/j.1365-246X.1968.tb00216.x}, making it numerically stable. We define $B[\vet{g}]$ to be dimensionless, namely
\begin{equation}\label{eq:B}
    B[\vet{g}] = \frac{E^2}{C_{ab}^{R}(a)^2}\; \sum_{\tau,\tau\p=1}^{\tau_{\mathrm{max}}} g_\tau \, \text{Cov}_{\tau \tau\p} \, g_{\tau\p} \; ,
\end{equation}
where $\text{Cov}$ is the covariance matrix of the correlator $C(a \tau)$ estimated over $N$ bins
\begin{equation}\label{covt}
    \text{Cov}_{\tau \tau\p}[C] = \frac{1}{N-1} \sum_{n=0}^{N-1} \left[ C_n(a\tau) - \braket{C(a\tau)} \right] \, \left[ C_n(a\tau\p) - \braket{C(a\tau\p)} \right] \; .
\end{equation}
The algorithmic parameters can be gathered to simplify the notation
\begin{equation}
    \vet{p} = (\alpha,\lambda, E_{\mathrm{min}}, \tau_{\mathrm{max}}) \; .
\end{equation}
The minimisation of $W_\alpha[\vet{g}]$ corresponds 
to solving the following linear problem,
\begin{equation}\label{eq:deW_deg}
    \frac{\delta  W_\alpha[\vet{g}]}{\delta g_\tau}\at_{\vet{g} = \vet{g}^{\vet{p}}} = 0 \; ,
\end{equation}
which has to be performed at each energy and smearing radius for which we want to estimate $\rho_{ab, \, \sigma}^{R}(E\p)$. 
The nature of the functionals appearing in these definitions is intimately related to the uncertainties on the estimator $\rho_{ab, \, \sigma}^{R}(E\p)$. The statistical error is in fact estimated by 
\begin{equation}\label{eq:stat_error_def}
    \Delta_{\mathrm{stat}}(E\p,\vet{g}^{\vet{p}}) = \frac{C_{ab}^{R}(1)}{E\p} \sqrt{ B[\vet{g}^{\vet{p}}]} \; .
\end{equation}
The systematic error, unavoidable at finite $\tau_{max}$, is estimated by monitoring the quantity 
\begin{equation}\label{eq:sys_error_def}
    d(\vet{g}^{\vet{p}}) = \sqrt{ \frac{A_0[\vet{g}^{\vet{p}}]}{A_0[\vet{0}]} } \; ,
\end{equation}
as we now describe. Regions where $d(\vet{g}^{\vet{p}})$ is small are dominated by the statistical uncertainty; on the other hand, the reconstructed smearing function of Eq.~\eqref{eq:kernal_bar} will be as close as possible to the exact one. For these reasons, it is not surprising that where $d(\vet{g}^{\vet{p}})$ is small, the results for $\rho_{ab, \, \sigma}^{R}$ are stable\footnote{Ref. \cite{Hansen_2019} shows that, with the value of $\tau_{\mathrm{max}}$ used in this work, we can expect a well reconstructed smearing kernel, 
at vanishing $\lambda$.} in response to variations of the algorithmical, unphysical parameters $\vet{p}$, within statistical error. If we define the coefficients $\vet{g}^{\star}$ as
\begin{equation}\label{eq:ra0}
    A_\alpha[\vet{g}^{\star}] / A_\alpha[\vet{0}] = k \;  B[\vet{g}^{\star}]\,   ,
\end{equation}
we find, with the given normalisations of the functionals and with the quality of our data, that in the range $0.1 < k < 5$ and $\alpha=1 $ the outcome of the reconstruction is in a region of algorithmical stability: $d(\vet{g}^{\vet{p}})$ is small, and systematic fluctuations are well within the statistical uncertainty. When this is not realised, the systematic error is estimated as
\begin{equation}\label{eq:sys_err_estimate}
    \Delta_{\mathrm{sys}}(E\p) = |\rho_{ab, \, \sigma}^{R}(E\p, \vet{g}^\star)- \rho_{ab, \, \sigma}^{R}(E\p, \vet{g}^{\star\star})| \; ,
\end{equation}
where $\vet{g^{\star \star}}$ is defined through Eq.~\eqref{eq:ra0} at a different value of $k$ that allows us to account for the systematic fluctuations. In this work, we find that a definition of $g^{\star \star}$ at $k/10$ provides a conservative estimate of $\Delta_{\mathrm{sys}}(E\p)$ for a reconstruction performed with $g^{\star}$ at the value $k$. Fig.~\ref{fig:lscan} shows an example, for a specific energy $E_\star$, of a stability regime where the spectral reconstruction for different values of $\lambda$ (black points) does not fluctuate outside the statistical error (black bars). Three points on the plot are highlighted: the ones corresponding to the choices of $k=0.1$, $k=2.5$ and $k=1$ in Eq.~\eqref{eq:ra0}, showing that indeed no systematic component affects the uncertainty on the result. The value corresponding to $k=1$ is associated to the value of $\lambda$ at which one achieves the optimal balance $ A_\alpha[\vet{g}^{\star}] / A_\alpha[\vet{0}] =\;  B[\vet{g}^{\star}]$, in agreement with the prescription from Refs.~\cite{Hansen_2019, Bulava:2021fre}.

\FloatBarrier

\subsection{Excited states in the antisymmetric sector}
\label{sec::measurements}

The excited states created by hadronic interpolators have a big impact on the extraction of the effective masses, since it can be hard to distinguish them from the ground state. In theories with multiple representations (or even just more flavors of a single representation) this effect is amplified when all the fermions are approaching the chiral limit. An interpolator can now create states containing particles from both representations. Since we simulate lighter fundamental fermions, we can expect this feature to be more visible in the 2AS sectors rather the in the fundamental one. Moreover, the 2AS sector does not have $G$-parity selection rules preventing certain states to mix, and we thus expect this channel to have a richer dynamics. In order to go into more details, it is useful to set the notation at least for the pseudoscalar interpolators
\begin{equation}
    \pi(x) = \bar{\psi}^{(\mathrm{F})}(x) \gamma_5 \psi^{(\mathrm{F)}}(x) \; , \;\;\;\;\; \Pi(x) = \bar{\psi}^{(\mathrm{2AS})}(x) \gamma_5 \psi^{(\mathrm{2AS})}(x) \; ,
\end{equation}
and the correlation functions
\begin{equation}
    C_{\mathrm{PP}}^{\mathrm{(F)}}(t) = \frac{1}{L^3} \sum_{\vet{x}} \braket{\pi(\vet{x},t) \bar{\pi}(0)} \; , \;\;\;\;\; C_{\mathrm{PP}}^{(\mathrm{2AS})}(t) = \frac{1}{L^3} \sum_{\vet{x}} \braket{\Pi(\vet{x},t) \bar{\Pi}(0)} \; .
\end{equation}
The importance of both identifying and controlling the excited states can be appreciated by looking at the smeared spectral density, for instance, in the pseudoscalar channel
\begin{equation}\label{eq:smeared_spectraldens_explicit_Again}
    \rho_{\sigma , PP}^{L,R}(E) = \sum_n \frac{\braket{0|O^{R}_\mathrm{P}(0)|n}_L \braket{n|\bar{O}^{R}_\mathrm{P}(0)|0}_L}{2E_n(L)}\, \Delta_\sigma \left( E-E_n(L) \right) \; .
\end{equation}
The magnitude of each matrix element $\braket{n|\bar{O}^{R}_\mathrm{P}(0)|0}_L$ determines the weight of each Gaussian $\Delta_\sigma$ located at the energy $E_n(L)$. If these energies are too close, or the matrix elements are too large, resolving different states can become laborious. With this motivation, in order to control the excited states we build correlation functions including two different types of fermionic field: local, point-like operators and Gaussian-smeared ones\footnote{Operator smearing is not to be confused with the smearing of spectral densities.}. Operator smearing allows working with interpolating operators that have a weaker overlap with excited states. Details concerning the measurements of the correlation functions and operator smearing can be found in Appendix \ref{app:sec:measure}.

\begin{figure}[!htb]
    \centering
    \includegraphics[width=0.69\textwidth]{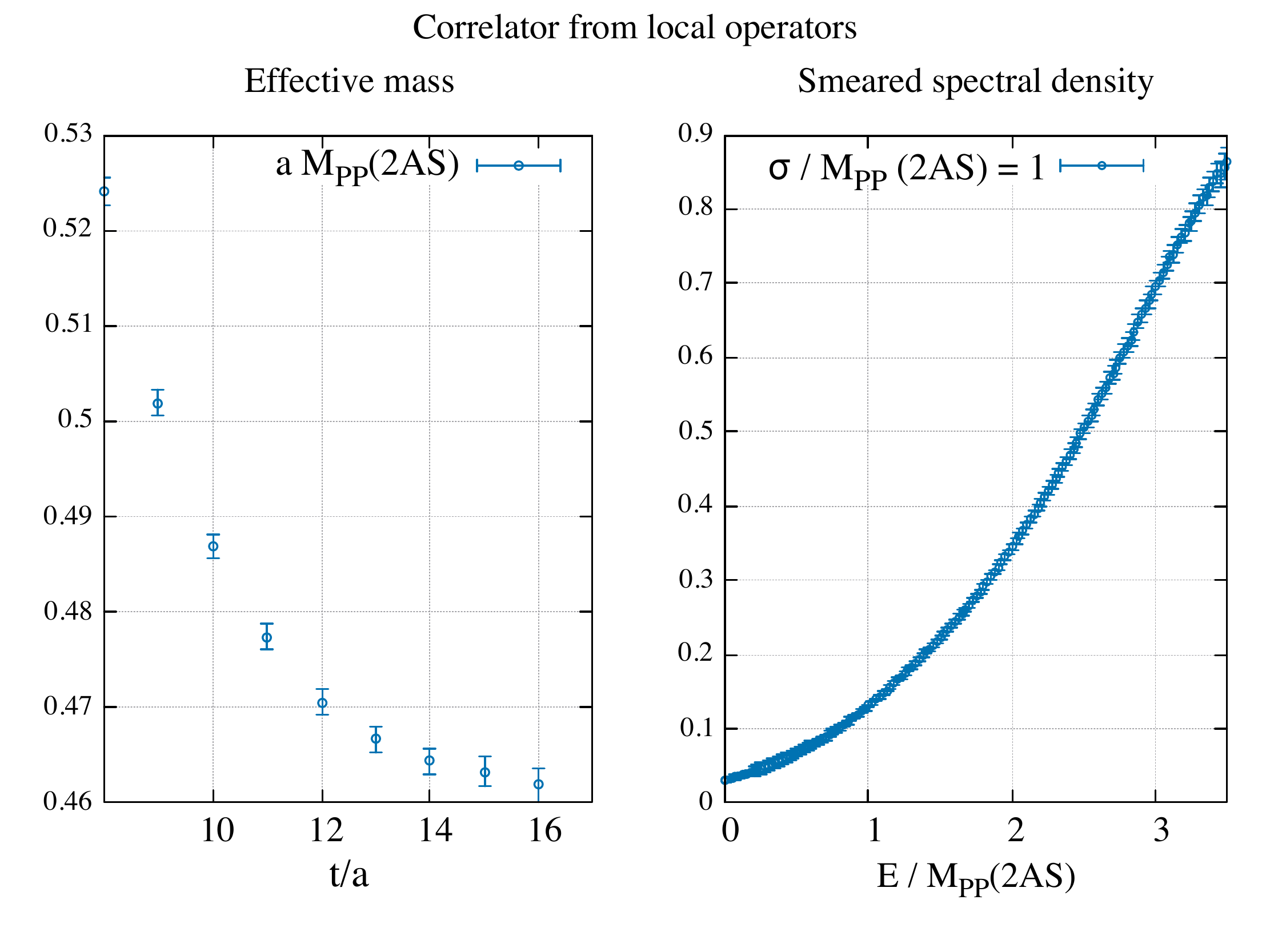}
    \caption{Results from a two point function of pseudoscalar operators built with point-like antisymmetric fermionic fields. The correlator is estimated from the ensemble B1. The left panel exhibits the effective mass as a function of time. Due to the nature of the excited states in the 2AS sector, the mass does not reach a plateau in the available time. The dominance of the excited states can also be understood from the smeared spectral density in the right panel. The overlap between the interpolator and the excited states it creates is too large: the spectral density smeared according to Eq.~\eqref{eq:gau_sm} is dominated by contributions above the multi-particle threshold, preventing the identification of the ground state.}
    \label{fig:signal_pointsource}
\end{figure}

The states created by each operator can be identified, as we have seen, by means of its symmetries. In the fundamental sector, at zero angular momentum, a pseudoscalar meson can induce the following transitions from the vacuum
\begin{equation}
    \braket{0|\pi|\pi} \; , \;\;\;\;\; \braket{0|\pi|\pi\pi\pi} \; , \;\;\; \braket{0|\pi|\pi\Pi\Pi} \; , \dots
\end{equation}
that will enter in our analysis through Eq.~\eqref{eq:smeared_spectraldens_explicit_Again}. Since in our simulations $M_{\mathrm{PP}}^{\mathrm{(2AS)}} > M_{\mathrm{PP}}^{\mathrm{(F)}}$ this phenomenology is reminiscent, up to $E_{3\pi}$,  of QCD, and one can expect computational aspects to be also similar. Conversely, due to the triviality of its $G$-parity, the 2AS sector has a multi-particle threshold located at $E_{2\Pi}$. In addition, since the other representation has lighter particles, states containing pseudoscalar mesons from the fundamental sector are not guaranteed to have energies far from the ground state. Possible overlaps with a pseudoscalar meson are in fact
\begin{equation}
    \braket{0|\Pi|\Pi} \; ,  \;\;\; \braket{0|\Pi|\Pi \,\Pi} , \;\;\; \braket{0|\Pi|\Pi \pi \pi} \, , \;\;\;  \braket{0|\Pi|\Pi \pi \pi \pi \pi} \, ,  \dots
\end{equation}
The aforementioned features complicate the extraction of $M_{\mathrm{PP}}^{\mathrm{(2AS)}}$, as it can be understood from Fig. \ref{fig:signal_pointsource} where we show two different types of signal from a correlator built with local, unsmeared operators. The excited states contaminate the signal for the ground state resulting, in the left panel, in an effective mass that does not reach a clear plateau. The problem is also manifest in the energy picture, as it is shown in the right panel of the same figure, where the smeared spectral density $\hat{\rho}_{\mathrm{PP}, \, \sigma}^{\mathrm{(2AS)}}$ does not exhibit the expected Gaussian peak around the mass of the pseudoscalar meson, but it rather grows monotonically. Indeed, by decreasing the smearing radius $\sigma$ of Eq.~\eqref{eq:gau_sm} one should be able to resolve such peak, but this cannot be realised with the current quality of the data. While the temporal length of the lattice poses an intrinsic obstacle to the thermalisation of the effective mass, the spectral reconstruction in principle allows obtaining smaller smearing radii also by increasing the number of configurations, since the systematic and the statistical error are related by Eq.~\eqref{eq:W_def}.

\FloatBarrier

\begin{figure}[htb]
    \centering
    \includegraphics[width=0.69\textwidth]{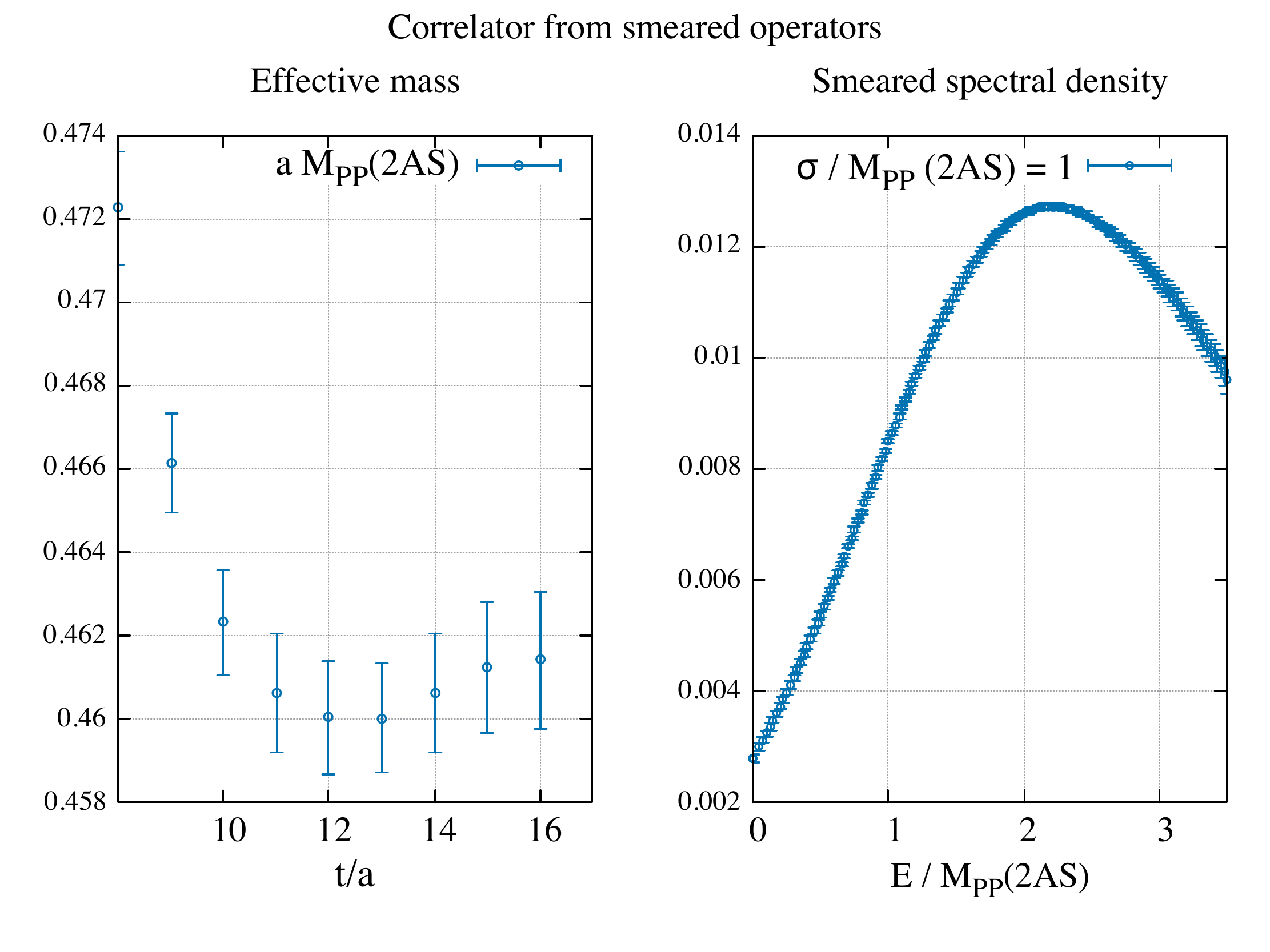}
    \caption{Results from Fig. \ref{fig:signal_pointsource}, this time using Gaussian-smeared interpolators according to Appendix \ref{app:sec:measure}. These operators are tuned to have smaller overlaps with the excited states. Consequently, the effective mass plot on the left reaches a plateau, providing an estimate for $aM_{\mathrm{PP}}^{\mathrm{(2AS)}}$. The right panel similarly shows how suppressed excited states allow for a clear peak to emerge in the spectral reconstruction smeared with $\sigma = M_{\mathrm{PP}}^{\mathrm{(2AS)}}$ according to Eq.~\eqref{eq:gau_sm}. The peak includes contributions from mainly $M_{\Pi}$ and $E_{\Pi\Pi}$.}
    \label{fig:signal_gaussiansm}
\end{figure}

Having established that the excited states present a challenge in the 2AS sector, it is natural to look at correlation functions of smeared operators defined in Appendix~\ref{app:sec:measure}, which have suppressed overlap with the excited states. 
Fig.~\ref{fig:signal_gaussiansm} shows the effective mass and the spectral reconstruction from the correlation function of such operators. On the left panel, the plateau in the effective mass shows an improvement compared to the corresponding result in 
Fig.~\ref{fig:signal_pointsource}. The effective mass is independent of time, within its statistical errors, for $t/a>10$. The smeared spectral density, shown in the right panel, demonstrates again the suppression of the excited states, with contributions from higher-energy states becoming smaller. As a result, a single peak is clearly visible at $E \simeq 2 M_{\mathrm{PP}}^{\mathrm{(2AS)}}$. The observed smeared spectral density is the result of two contributions coming mainly from the energy levels $M_{\mathrm{PP}}^{\mathrm{(2AS)}}$ and $E_{\Pi \Pi}\simeq 2 M_{\mathrm{PP}}^{\mathrm{(2AS)}}$, which cannot be resolved because the smearing radius is too large, $\sigma \simeq M_{\mathrm{PP}}^{\mathrm{(2AS)}}$. These energies can be nonetheless estimated by fitting the spectral density to a sum of Gaussians. This idea will be expanded in the next section.

\FloatBarrier

\subsection{Fits of spectral densities}
\label{sec:spectre}

In this section we describe fit strategies for spectral densities. A parallel discussion on fits of correlators will highlight the differences between the two methodologies and will lead to a quantitative comparison between predictions for the pseudoscalar masses obtained from the two approaches, which is presented in detail at the end of the section. The model functions used in the fits are
$c^{(k)}(t)$ for correlators and $f_
\sigma^{(k)}(t)$ for the smeared spectral densities
\begin{equation}
\label{eq:fitmodels}
    g^{(k)}(t) = \sum_{n=1}^k a_n \left( e^{-t E_n} + e^{(-T+t)E_n} \right)  \; , \;\;\;\;\; f^{(k)}_\sigma(E) = \sum_{n=1}^k b_n e^{-(E-E_k)^2/\, 2\sigma^2} \; ,
\end{equation}
where $\sigma$ is the smearing radius defined by Eq.~\eqref{eq:gau_sm}. The integer $k$ encodes how many states are included 
in our model function. $E_n$, $a_n$ and $b_n$ are the fit parameters which relate to finite volume energies and matrix elements. These are estimated by minimising appropriate $\chi^2$ functions
\begin{align}
\label{eq:chisq_T}
    \chi_{g^{(k)}}^2 &= 
    \sum_{t, t'} \left( g^{(k)}(t) - C(t) \right) 
    \text{Cov}^{-1}_{tt'}[C] \left( g^{(k)}(t') - C(t') \right) \; , \\
\label{eq:chisq_E}
    \chi_{f_\sigma^{(k)}}^2 &= 
    \sum_{E,E'} \left( f_\sigma^{(k)}(E) - \rho_\sigma(E) \right)
    \text{Cov}^{-1}_{EE'}[\rho_\sigma]\left( f_\sigma^{(k)}(E') - 
    \rho_\sigma(E') \right)\; ,
\end{align}
where covariance matrices are estimated as in Eq.~\eqref{covt} both for correlators and spectral densities.

On the lattice, the temporal length $T$ constrains the maximum number of data points and hence degrees of freedom for fitting a correlator $C_{\mathrm{PP}}^{R}(t)$. 
The {\em effective} number of degrees of freedom
may be further constrained in the presence of correlated data. The effect of correlation between times $t$ and $t'$ is taken into account by the covariance matrix appearing in $\chi_{g^{(k)}}^2$. 
A smeared spectral density $\rho^{R}_{\mathrm{PP},\, \sigma}(E)$ can be in principle evaluated for any energy $E$, yet being it derived from the correlator itself, it seems unrealistic to think it could lead to a larger number of degrees of freedom to be used in a fit. The information, however, is mixed non-trivially: the spectral density at a single point depends on the correlator at each lattice time $t$. A consequence of this feature is manifested in the extraction of the ground state, where the problem of having uncontaminated signal at large times is shifted to having a small enough resolution $\sigma$ in the energy.

\begin{figure}[t]
    \centering
    \includegraphics[width=0.4\textwidth]{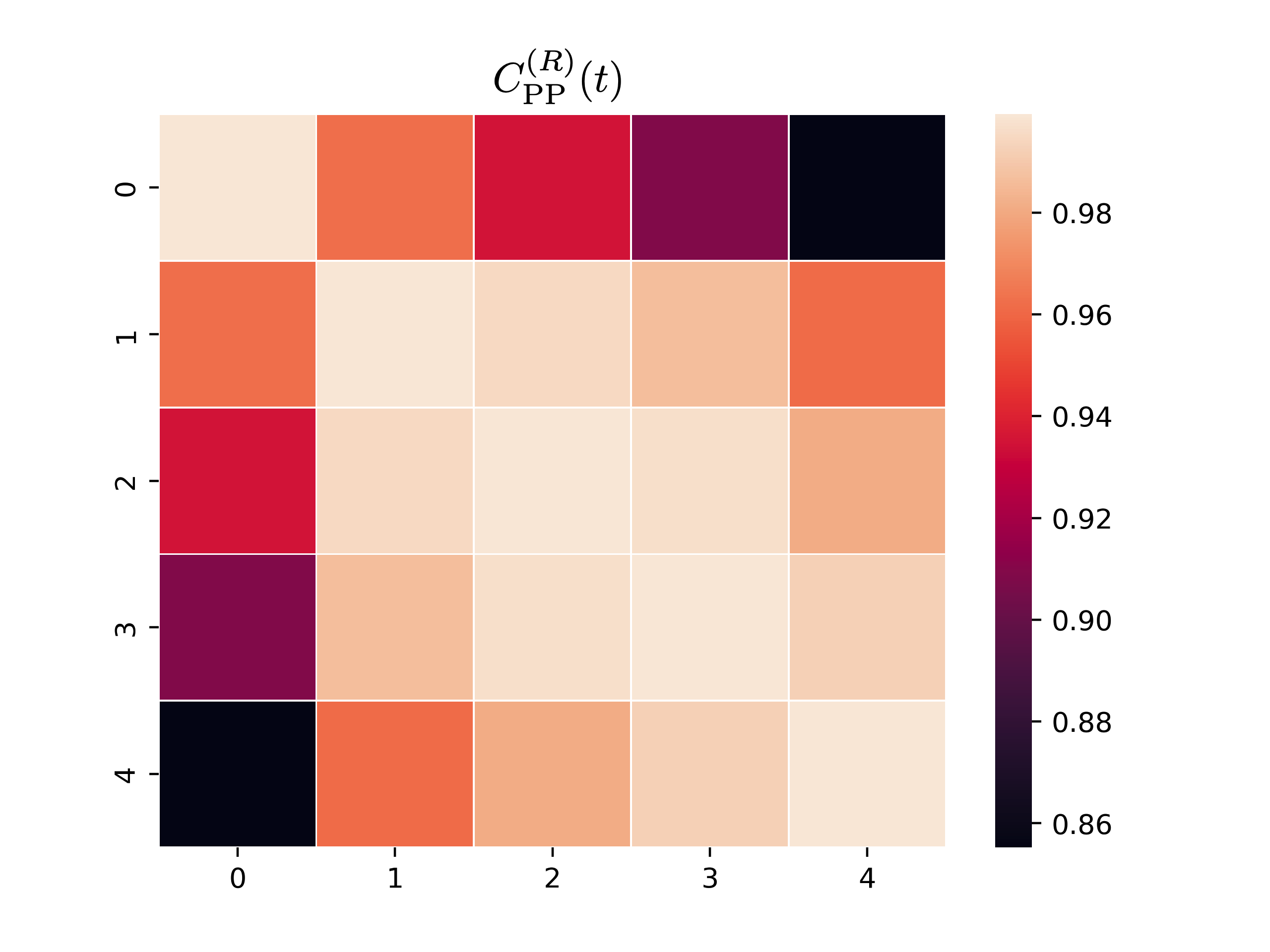}
    \includegraphics[width=0.4\textwidth]{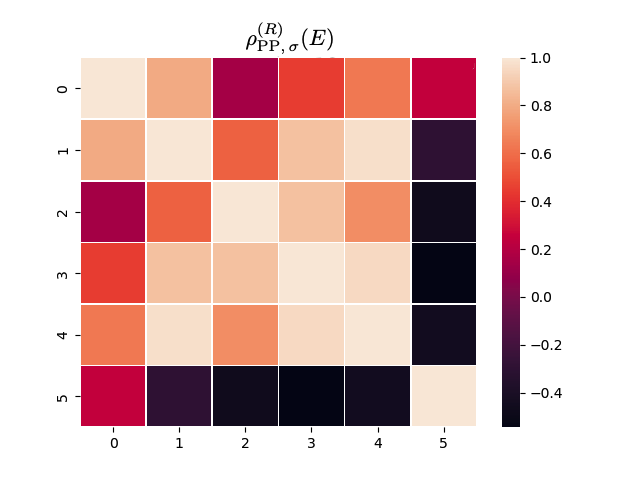}
    \caption{Covariance matrices for the lattice correlator $C_{\mathrm{PP}}^{(2AS)}(t)$ at five time slices (left) and the smeared spectral density $\rho_{\mathrm{PP}, \, \sigma}^{\mathrm{(2AS)}}(E)$ evaluated at six energies (right) from $C_{\mathrm{PP}}^{\mathrm{(2AS)}}(t)$. The points at which the spectral density is evaluated are chosen in order to minimise the condition number of its covariance matrix. Due to this freedom, we obtain a matrix for the spectral density that is better conditioned than the one for the correlator.}
    \label{fig:covariance_matrices}
\end{figure}

With this motivation, it is interesting to study the number of degrees of freedom we can exploit in each correlated fit. While the correlators can only be evaluated at integers $0\leq t < T$, we have freedom to choose the energies in a given interval at which we evaluate the spectral densities. Our criterion is to select those that minimise the condition number of $\text{Cov}[\rho_\sigma]$ in order to maximise the information that is passed to the $\chi^2$. We observe that correlated data in time does not necessarily translate into correlated data in energy space, as shown in Fig.~\ref{fig:covariance_matrices}, where we compare covariance matrices for $C_{\mathrm{PP}}^{(2AS)}(t)$ and $\rho_{\mathrm{PP}, \, \sigma}^{\mathrm{(2AS)}}(E)$ from the ensemble B3 with a smearing radius $\sigma=0.2/a$. 

In both cases, the degrees of freedom are enough to fit at least two states. The matrix $ \text{Cov}^{-1}_{t t\p}[C]$, in particular, is evaluated from correlators measured from $t/a=8$ to $t/a=16$ with an interval of two. Regardless of the thinning in time, adjacent points shows a very high correlation. If the covariance matrix of the correlator $\text{Cov}[C]$ is ill-conditioned, it needs to be regularised by applying a cutoff on its smaller eigenvalues before $\chi^2_{g^{k}}$ is evaluated, a problem that is not faced when fitting spectral densities, whose covariance matrix is easier to invert. Indeed, one has to invert the covariance of the correlators in order to compute the spectral density, but only in the combination defined by Eq.~\eqref{eq:W_def}: the matrices obtained from the functionals $A_\alpha[\vet{g}]$ and $B[\vet{g}]$, if both ill conditioned, regularise each other for suitable values of $\lambda$. The choice of a cutoff for the covariance matrix $\text{Cov}[C]$ is therefore absorbed into the choice of the parameter $\lambda$.

\begin{figure}[t]
    \centering
    \includegraphics[width=0.49\textwidth]{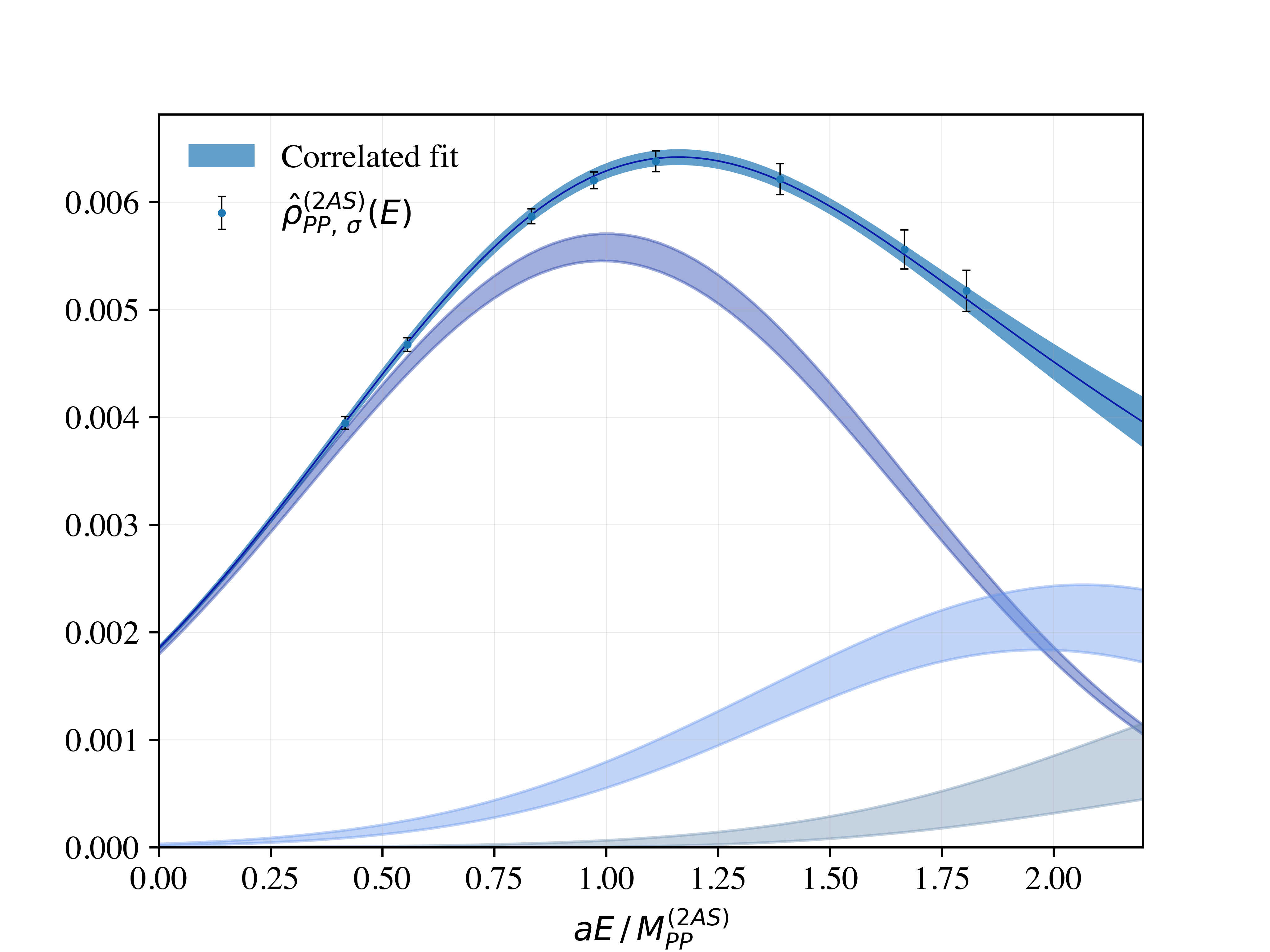}
    \includegraphics[width=0.49\textwidth]{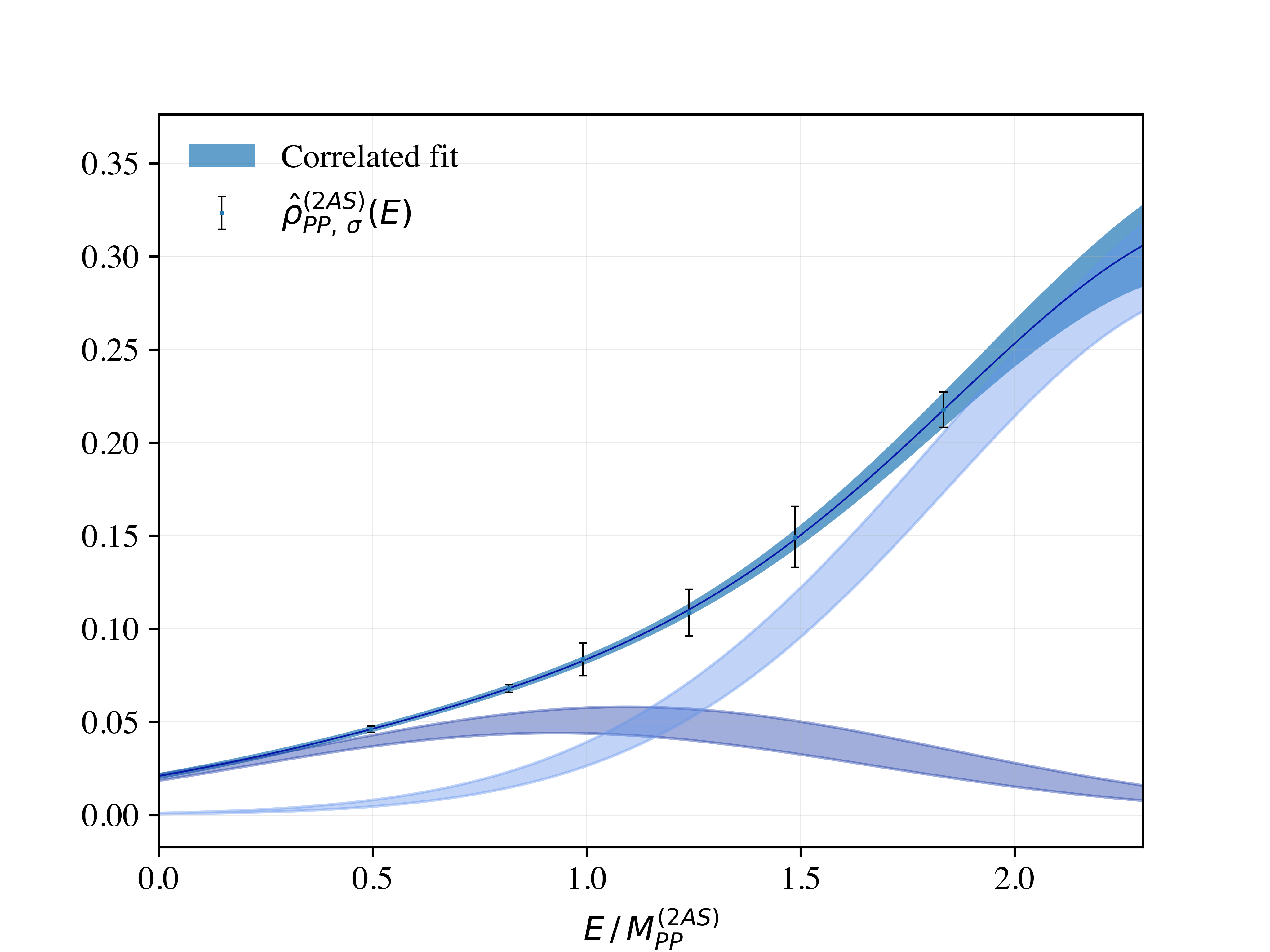}
    \caption{Examples of fits of spectral densities, showing the breakdown of the contribution of each Gaussian. On the left-panel, a three Gaussian fit of a smeared spectral density extracted from a correlator that uses smeared, non-local fields. Due to this choice for the interpolator, the Gaussians in the plot contribute less as we go higher in the energy range. On the right-panel, we show a similar plot obtained from local interpolators: the two Gaussian fit is still able to isolate the ground state even if the effective mass does not plateau. Both plots correspond to the pseudoscalar 2AS channel. The left panel is obtained within the ensemble B3 and the spectral density has smearing radius $\sigma=0.24/a$. The right panel is derived from the ensemble B2 and the smearing radius of the spectral density is $\sigma=0.3/a$.}
    \label{fig:3rho}
\end{figure}

Fig.~\ref{fig:3rho} shows two examples of correlated fits of the smeared spectral densities. On the left panel, the correlator used to extract $\rho_{\mathrm{PP}, \, \sigma}^{(\mathrm{2AS})}$ is built with smeared interpolating fields. The model function is $f_\sigma^{(3)}$ from Eq.~\eqref{eq:fitmodels}. The plot shows the explicit breakdown of each Gaussian, which correspond to the contribution of different energy levels: as each of them becomes decreasingly important, we clearly see the excited state suppression achieved by the choice of smeared operators in the correlation function. At low energies, the spectral density is almost entirely dominated by the first energy level, therefore the corresponding fit parameter is mainly constrained by energies near the origin. Cancellations between the three correlated contributions combine in an error on the fit result that is generally smaller than the one on the single Gaussians. We identify the first peak as the value of $aM_{\mathrm{PP}}^\mathrm{(2AS)}$ and the second with $a E_{\Pi\Pi}$. The right panel of Fig.~\ref{fig:3rho} is instead obtained from local interpolators, which have a larger overlap with excited states. As shown in Fig.~\ref{fig:signal_pointsource}, the effective mass of Eq.~\eqref{eq:cosh_mass} does not reach a clear plateau in this case, yet the fit of the spectral density is able to isolate the ground state. The error on the fit, however, is at best one order of magnitude larger than the corresponding one obtained in the left panel. The choice of smeared operators is therefore preferred. 

\begin{figure}[t]
    \centering
    \includegraphics[width=0.69\textwidth]{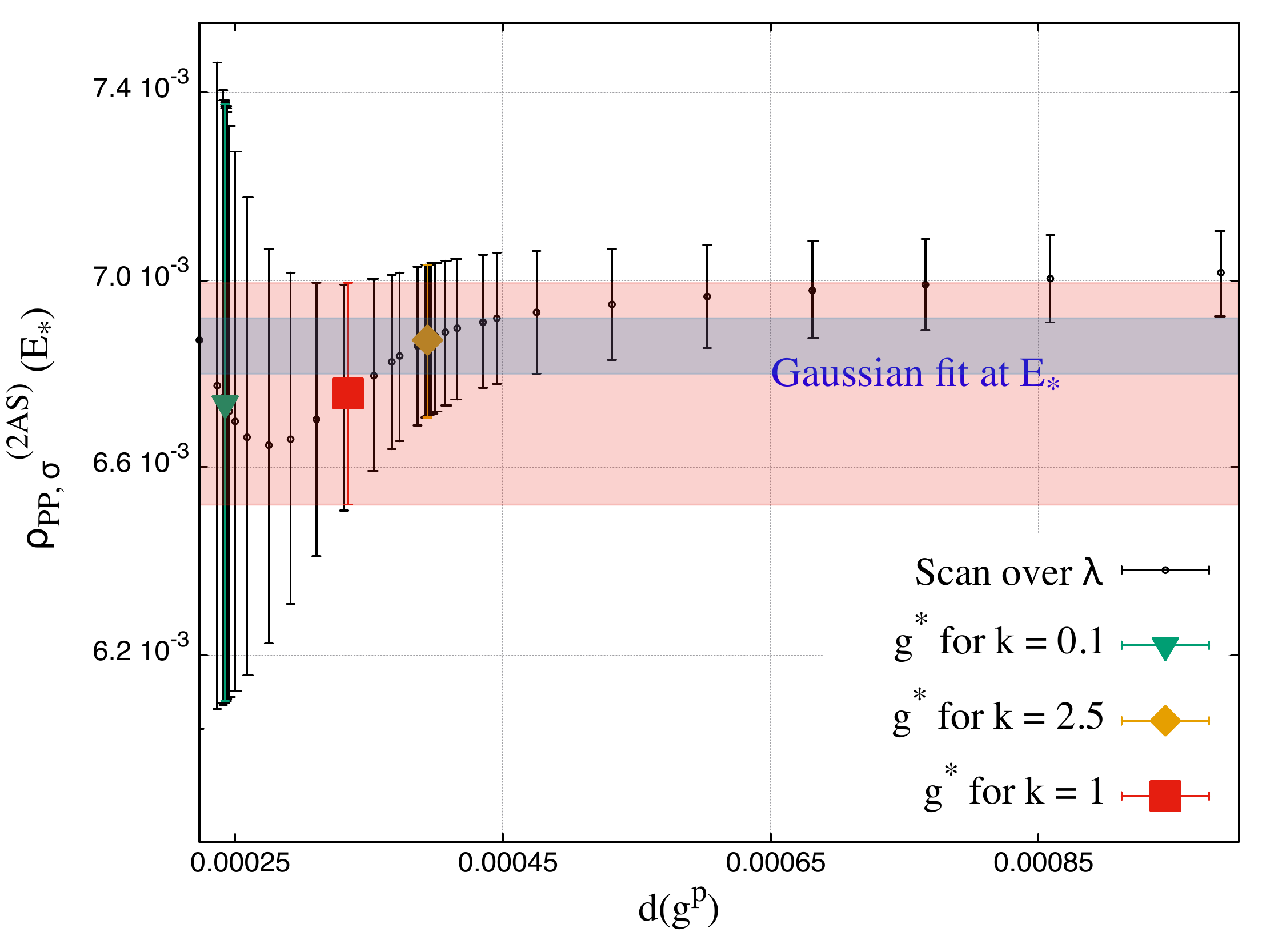}
    \caption{The plot updates Fig. \ref{fig:lscan}, which shows that the reconstruction (red band) does not change outside the statistical error (black bars) for different choices of the unphysical parameter $\lambda$, in the given range of $d(\vet{g}^{\vet{p}})$ (cf. Eq.~\eqref{eq:sys_error_def}). The blue band is the fit result of the smeared spectral density to a sum of Gaussians at the point $E_\star$. Encouragingly, the fit is compatible with all points in the scan, showing that stability in the reconstruction translates into stability for the fits.}
    \label{fig:lambdascan_wfit}
\end{figure}

We now turn to the comparison of the fit results. As described in Sec.~\ref{sec::measurements}, the extraction of the ground state presents challenges in the antisymmetric sector, which therefore provides
an interesting testbed to compare the two frameworks. We use for the comparison the ensembles B1-B4, where both representations are fairly light. We begin by discussing the estimate of $a M_{\mathrm{PP}}^{\mathrm{(2AS)}}$ from lattice correlators. In some cases, the covariance matrix of the two point function had to be regularised by introducing a cutoff on its lower eigenvalues, in order to invert it in the $\chi^2$. The choice of this cutoff can translate into fluctuations in the estimate of the pseudoscalar mass outside the statistical error, that have been accounted for by adding a systematic component to the uncertainty. This problem does not appear while fitting spectral densities, because the covariance of the spectral density is better conditioned. We have also ensured that no contamination was present in the estimate for the pseudoscalar mass by comparing fits of one and more exponentials. The values of the pseudoscalar mass obtained with this approach can be found in Table \ref{tab:PSmesons}. In this framework the smearing of the interpolators has been crucial, as it is clear from Fig.~\ref{fig:signal_pointsource} where the effective mass of Eq.~\eqref{eq:cosh_mass} does not plateau due to the short temporal extent of the lattice. The spectral density, on the other hand, does not rely on any large time behaviour. Its limit lies in the high energy range, where its error becomes large. In order to perform the reconstruction, the algorithmic inputs are chosen in the region of stability as described in Sec.~\ref{sec:backus-gilbert}. 
Fig.~\ref{fig:lambdascan_wfit} updates the plot from Fig.~\ref{fig:lscan} showing that stability for the reconstruction translates into stability for the fits: the blue band, which is the fit result for the smeared spectral density at the energy $E_\star$, is compatible with all the values generated by different choices for the unphysical parameter $\lambda$.

\begin{figure}[b!]
    \centering
    \includegraphics[width=0.69\textwidth]{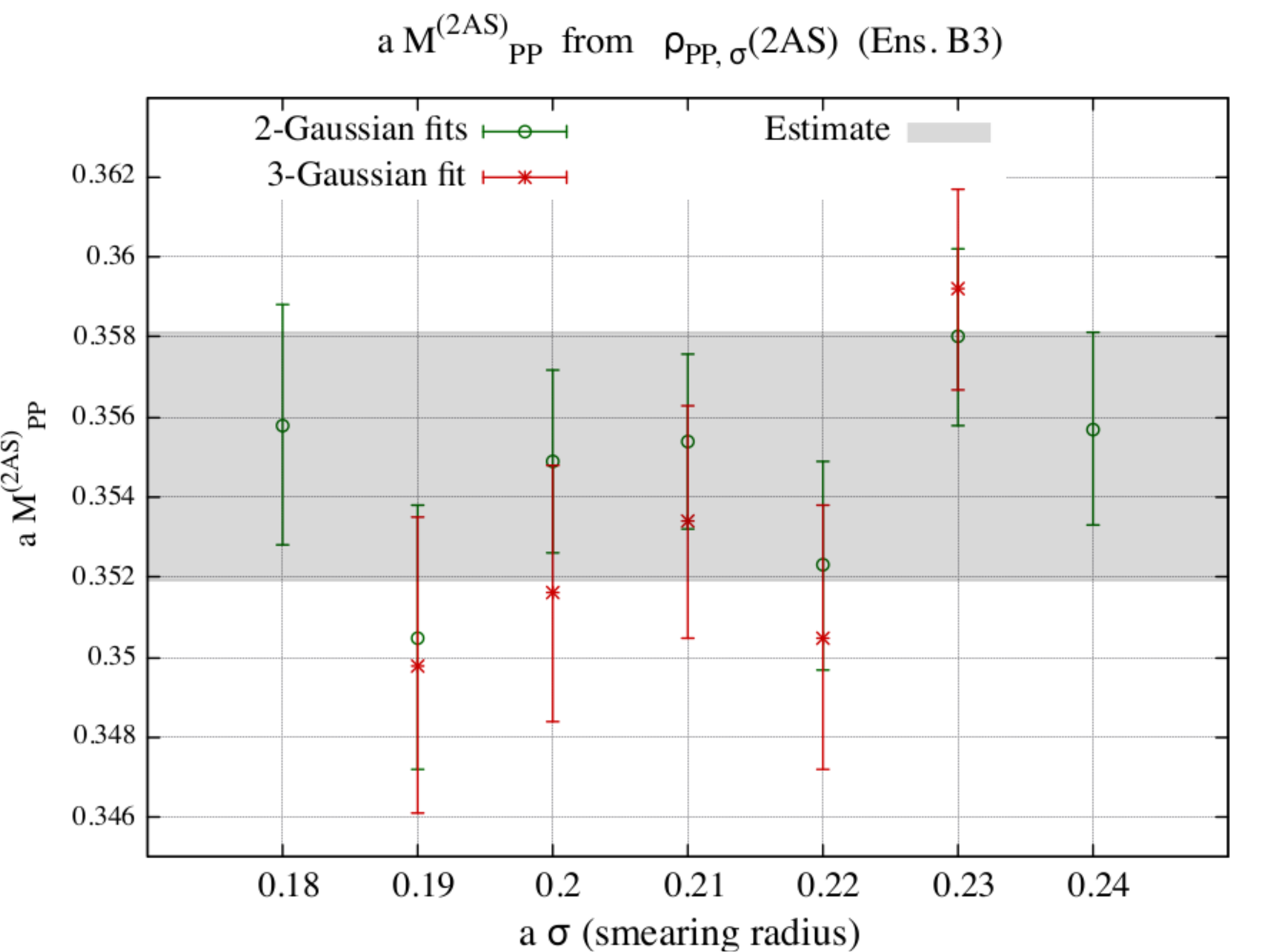}
    \caption{Fit results for $a M_{\mathrm{PP}}^{(\mathrm{2AS})}$ from the ensemble B3, obtained from two (green) and three (red) Gaussian fits of smeared spectral densities at different smearing radii $\sigma$. Fluctuation at different values of the smearing radius translate into a systematic component of the uncertainty. This is summed in quadrature to the statistical error in the gray, horizontal band, the estimate for the pseudoscalar mass $a M_{\mathrm{PP}}^{\mathrm{(2AS)}} = 0.3550(31)$.}
    \label{fig:finalres}
\end{figure}

The choice of the smearing radius is dictated by the quality of the data. In this work, we managed to obtain values ranging from $\sigma=0.18/a$ to $0.33/a$. Since the separation between finite volume energies should be roughly  $2\pi/L \simeq 0.4$, these values can be considered acceptable. For each ensemble, we have performed the fit at a fixed value of $\sigma$, obtaining a prediction for the pseudoscalar mass  $aM_{\mathrm{PP,\, \sigma}}^{(\mathrm{2AS})}$. We then performed a scan over different smearing radii to check for systematic effects. As shown in the example of Fig. \ref{fig:finalres}, the smearing radii adopted were found to be small enough to identify the ground state; in most cases, still, we have observed fluctuations for $aM_{\mathrm{PP,\, \sigma}}^{(\mathrm{2AS})}$ as $\sigma$ varies. When they occurred, we added half the spread of these fluctuations as a systematic error. 
Fig.~\ref{fig:finalres} also shows a comparison between fits that include two and three states: the results are in good agreement, signalling that no contamination from excited states is affecting the estimate of the mass. 

\begin{figure}[ht]
    \centering
    \includegraphics[width=0.41\textwidth]{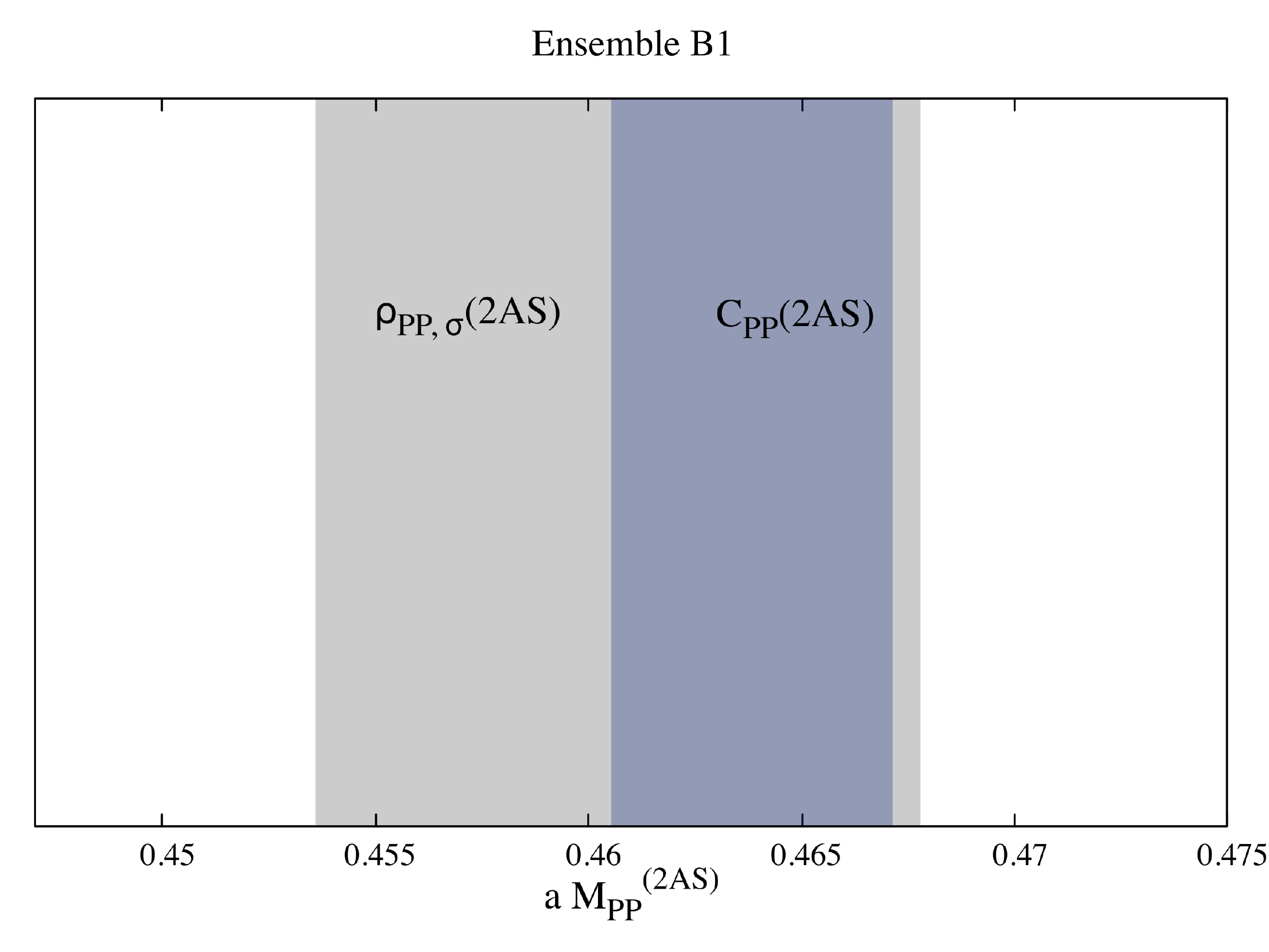}
    \includegraphics[width=0.41\textwidth]{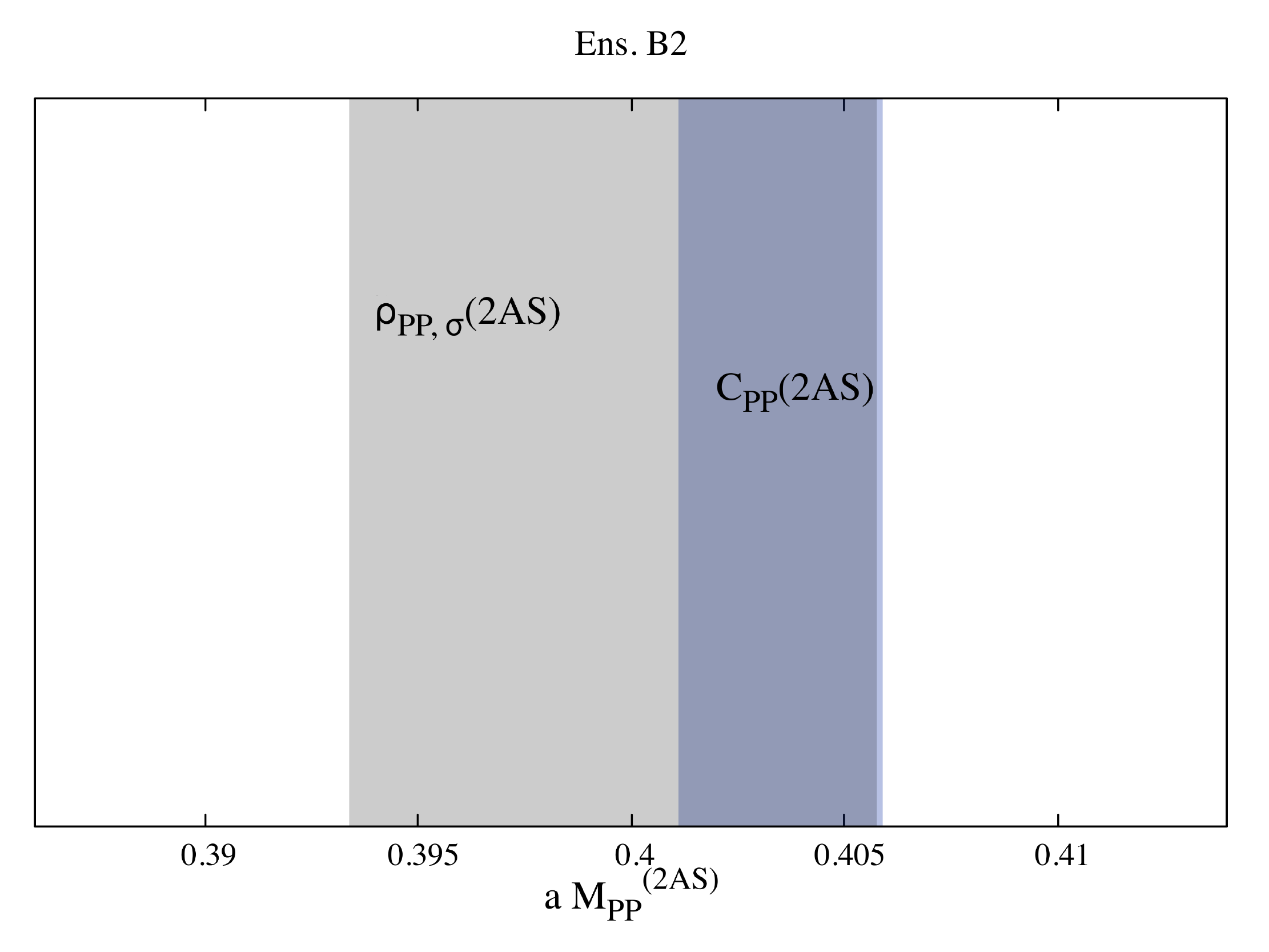}
    \includegraphics[width=0.41\textwidth]{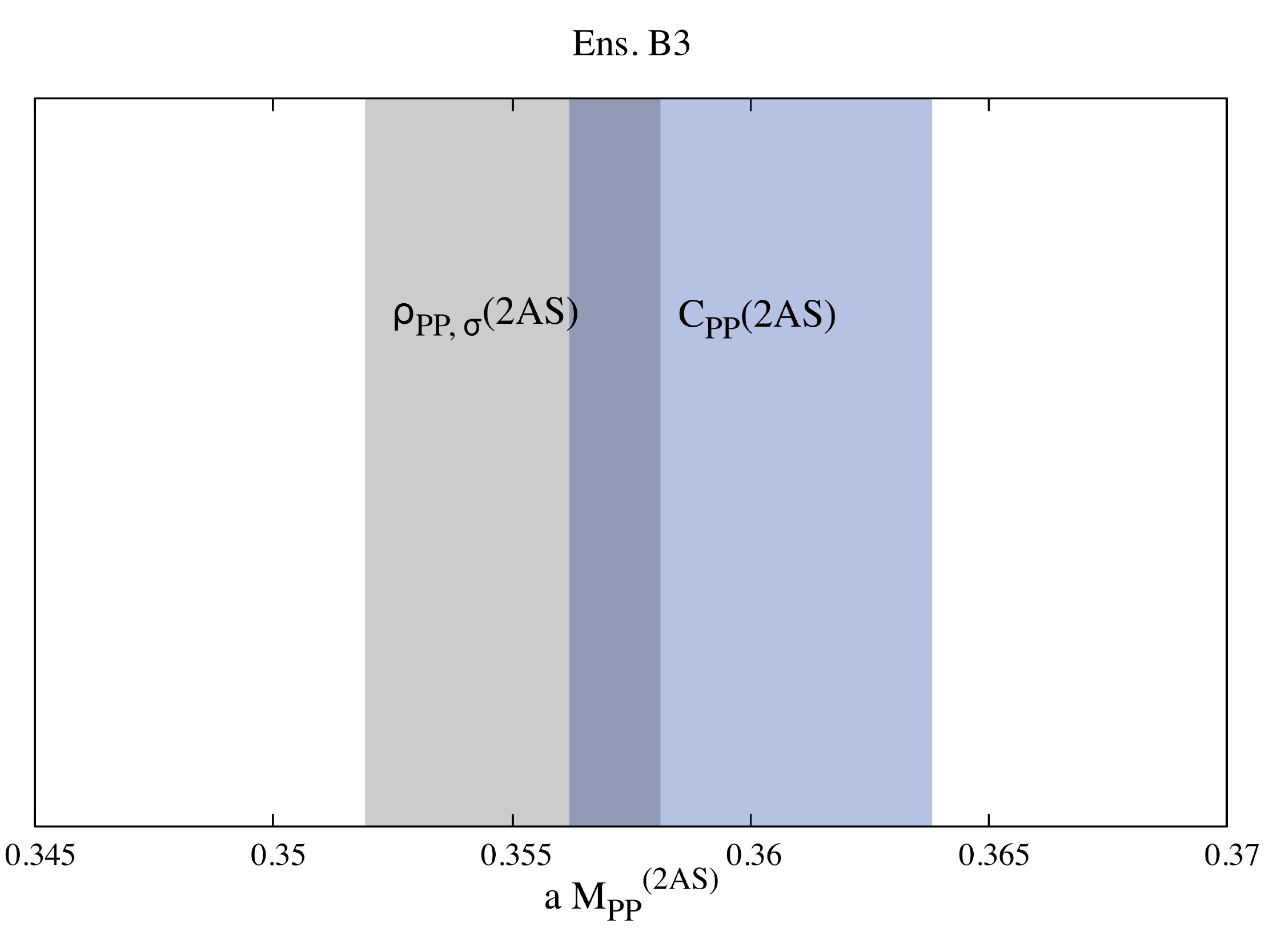}
    \includegraphics[width=0.41\textwidth]{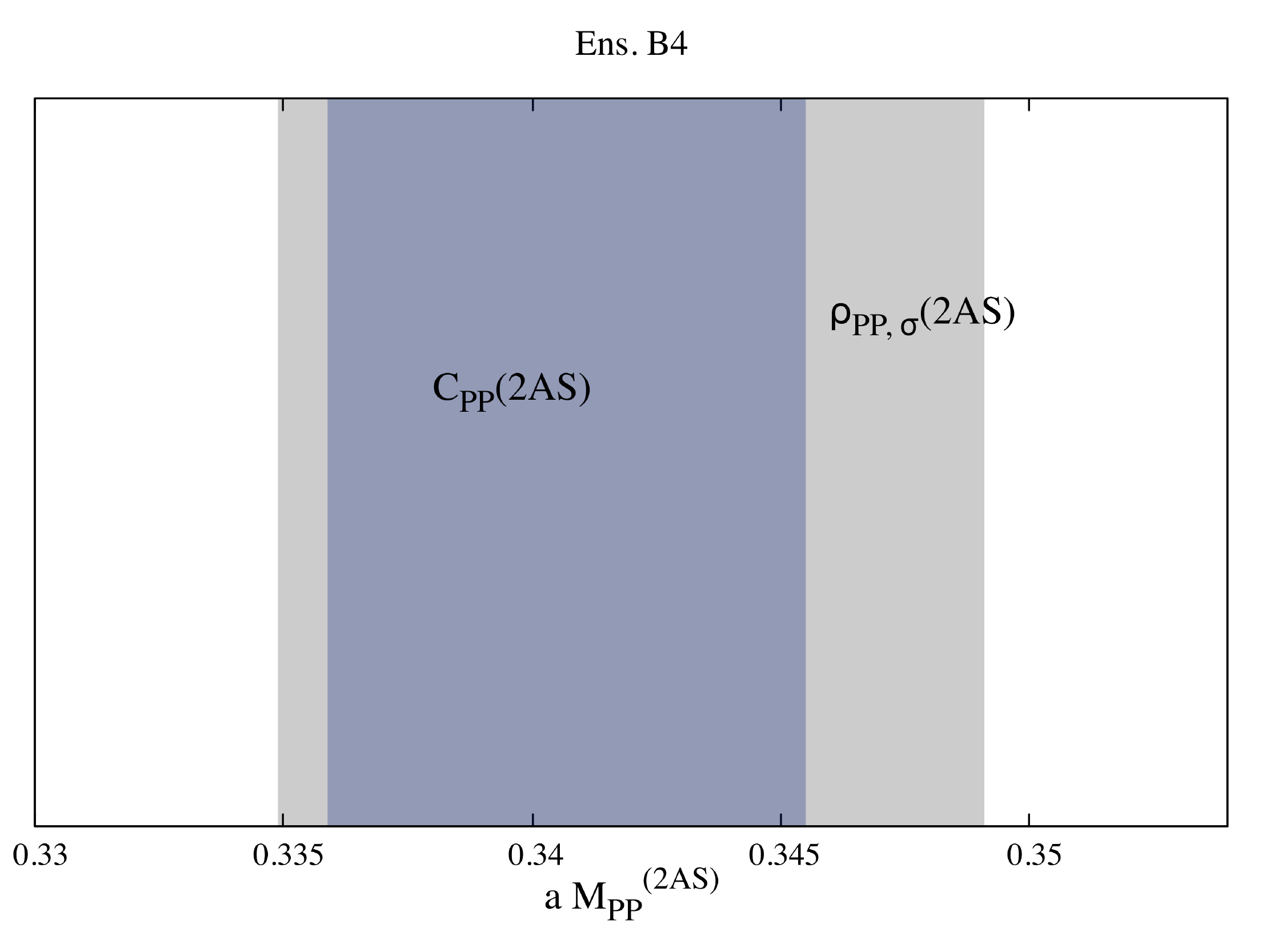}
    \caption{Graphical comparison between the two predictions for $a M_{\mathrm{PP}}^{(\mathrm{2AS})}$ for the ensembles B1-B4.}
    \label{fig:cmp}
\end{figure}

The comparison between fits of spectral densities and correlators is shown in Fig.~\ref{fig:cmp}. The predictions are always compatible, and the errors lie in the same order of magnitude, but the uncertainty on the correlator is generally smaller, up to a factor of approximately two. 

The numerical values used in this comparison are listed in Table \ref{tab:cmp}. It should be noticed that the outcome of this comparison holds with the given amount of statistics and time extent of the lattice. These quantities, in fact, heavily influence the analysis both on the side of the correlators and the spectral density, yet the way in which the two methodologies are affected can be different.
\begin{table}[h!]
    \centering
    \begin{tabular}{c | c | c }
        Ensemble & $a M_{\mathrm{PP}}^{\mathrm{(2AS)}}$ from $C_{\mathrm{PP}}^{(2AS)}(t)$ & $a M_{\mathrm{PP, \sigma}}^{\mathrm{(2AS)}}$ from $\rho_{\mathrm{PP}, \, \sigma}^{\mathrm{(2AS)}}(E)$ \\[-1em]\\ \hline
          \\[-0.8em]
         B1 & 0.4638(33) & 0.4607(71)\\
         B2 & 0.4035(24) & 0.3996(88) \\
         B3 & 0.3600(38) & 0.3550(31)\\
         B4 & 0.3407(48) & 0.3420(71)\\
    \end{tabular}
    \caption{Predictions for the pseudoscalar mass in the 2AS sector from different ensembles. The values are depicted in Fig. \ref{fig:cmp}.}
    \label{tab:cmp}
\end{table}
\FloatBarrier

\section{Conclusions}
\label{sec:z_conclusions}

The model studied in this work allows us
to understand the dynamics of partial compositeness better, by  extending the effort started in Ref.~\cite{Cossu_2019} at a reasonable computational cost. In order to make contact with the phenomenology of these theories from a lattice perspective, the systematics related to the computation of the spectrum and to the extrapolation to the chiral limit need to be under control. Our analysis has been developed in this context. 
We have explored the perturbative structure of the theory, extending previous computations of the critical mass~\cite{Follana:2000mn, Panagopoulos:1998xf, DelDebbio:2008wb, PhysRevD.86.014505, PhysRevD.74.074503} to the case of multiple fermionic representations. We have computed the self-energy of a fermion in a given representation, focusing on the dynamical effects due to the presence of more fermionic fields in a different representation of the gauge group. 
While leaving a clear footprint, the second representation has only a minor numerical impact on the value of the critical mass. 
This result, while perturbative, found a counterpart at the non-perturbative level in the outcome of our simulations, as it is shown in Tables~\ref{tab:pcac_fund}--\ref{tab:PSmesons}. By approaching the chiral limit in a given representation, in fact, we found a weak dependence on the bare parameters of the other one. The extrapolation to the chiral limit was based on several ensembles that were generated for this study. This task was hindered by a strong autocorrelation affecting observables built from the fermion fields in the fundamental representation, the lightest in terms of pseudoscalar masses. A great portion of this work focused on the extraction of the spectrum, a crucial task in order to understand if these models are realistic from a phenomenological perspective. We have shown that the 2AS sector is characterised by complicated dynamics, due to the interplay between different representations, and the lack of selection rules dictated by discrete symmetries. We have found operator smearing to be essential, in this context, to provide a reliable analysis of the lattice data. In our analysis, we have taken advantage of recent progresses \cite{Hansen:2017mnd, Bulava:2019kbi,  Bulava:2021fre, Gambino:2022dvu, Bruno:2020kyl, Bailas:2020qmv} in the numerical solution of the inverse problem that allow precise reconstructions of smeared spectral densities. Our approach, introduced in Ref.~\cite{Hansen_2019}, yields spectral densities smeared with a chosen function and with controlled systematics. Due to these features it was possible, for the first time, to explore in this work the extraction of finite volume energies from fits of spectral densities. Our results, compared to other established approaches, provide complementary insights and fully compatible results. 

Encouraged by these conclusions, we plan to explore the possibility of extending this computational setup to the study of baryonic operators, as well as to other theories of composite Higgs. Our choices, both in terms of software \cite{boyle2015grid, antonin_portelli_2022_6382460} and computational framework, are general and easy to adapt for the study of other theories, whether they contain different number of colors, fermionic content or gauge groups. Moreover, our work clarified technical aspects of the partial compositeness dynamics: this important 
step will allow moving towards studies of more phenomenological relevance. Theories of partial compositeness provide a rich set of new particles, from pseudo Goldstone bosons to heavy-quark partners, that are charged under the Standard Model and could be important for direct and indirect search of new physics. The extraction of spectral densities, validated in this context by our work, can be used not only to extract energy levels but also to directly compute inclusive cross-sections~\cite{Gambino:2020crt, Hansen:2017mnd}, a possibility that we leave for future studies. The knowledge of inclusive processes can be important for the indirect search of new physics, with particles from the new sector leaving footprints in observables precisely measured both at present and future colliders.

\section*{Acknowledgements}
\label{sec:z_thanks}

AL and LDD received funding from the European Research Council (ERC) under the European Union’s Horizon 2020 research and innovation programme under grant agreement No 813942.  LDD is supported by the UK Science and Technology Facility Council (STFC) grant ST/P000630/1. The numerical simulations were run on machines of the Consorzio Interuniversitario per il Calcolo Automatico dell’Italia Nord Orientale (CINECA). We acknowledge support from the SFT Scientific Initiative of INFN.

\appendix

\section{Group-Theoretical Conventions}
\label{app:groups}

We denote the generators of $SU(N)$, the special unitary group of degree $N$, as $T^a$, with $1 \le a \le N^2-1$. They are traceless Hermitian matrices satisfying the algebra
\begin{equation}
    \left[T^a,T^b\right] = i \sum_{c=1}^{N^2-1} f^{abc} T^c \; ,
 \end{equation}
where $f^{abc}$ are the structure constants of the $A_{N-1}$ Lie algebra, which are antisymmetric under permutations of every pair of indices, and are the same in all representations. The generators in an irreducible representation $R$, denoted as $T^a_R$, satisfy
 \begin{equation}
 \label{eq:Dynkin_index}
\Tr \left( T_R^a T_R^b \right) = \lambda_R \delta^{ab} \; ,
 \end{equation}
 where $\lambda_R$ is the Dynkin index of the representation $R$. In our normalisations, the Dynkin index of the fundamental representation (fund) is $\lambda_{\mbox{\tiny{Fund}}}=\frac{1}{2}$. Another interesting group-theoretical invariant is the sum of the squares of the generators, which is the quadratic Casimir operator, and is proportional to the identity matrix in every representation; we denote the corresponding eigenvalue in the representation $R$ as $C_2(R)$:
 \begin{equation}
 \label{eq:quadratic_Casimir}
     \sum_{a=1}^{N^2-1} T_R^a T_R^a = C_2(R) \ide \; .
 \end{equation}
 Comparing Eq.~(\ref{eq:Dynkin_index}) and Eq.~(\ref{eq:quadratic_Casimir}) one obtains
 \begin{equation}
     \lambda_R = \frac{\text{dim}(R)}{N^2-1} \, C_2(R) \; ,
 \end{equation}
 $\text{dim}(R)$ being the dimension of the representation $R$. With these conventions, the values of these invariants for the fundamental and 2AS representation are listed in Table~\ref{tab:grouptheoryfactors}.
 \begin{table}[h!]
     \centering
     \begin{tabular}{c | c c c}
          $R$ &  $\text{dim}(R)$ & $\lambda_R$ & $C_2(R)$ \\[-1em]\\ \hline
          \\[-0.8em]
          Fund & $N$ & $\frac{1}{2}$ & $\frac{N^2-1}{2N}$ \\
          \\[-0.8em]
          2AS & $\frac{N(N-1)}{2}$ & $\frac{N-2}{2}$ & $C_2($Fund$) \frac{2(N-2)}{N-1}$ \\[-1em]\\ \hline
     \end{tabular}
     \caption{Group-theoretical factors used in this work.}
     \label{tab:grouptheoryfactors} 
 \end{table}
 
The explicit form of the generators that were used in this work is as follows. For the fundamental representation, the $N-1$ Cartan generators are defined as
\begin{equation}
\label{eq:fundamental_Cartan_generators}
     T_{\mbox{\tiny{Fund}}}^k = \frac{1}{\sqrt{2k(k+1)}} \,\mathrm{diag} \left( \underbrace{1,1,\ldots ,1,1}_{\mbox{\footnotesize{$k$ terms}}}, -k,\underbrace{0,0,0,\ldots,0,0}_{\mbox{\footnotesize{$N-k-1$ terms}}} \right) , \qquad \mbox{for $1 \le k \le N-1$},
\end{equation}
whereas the remaining $N^2-N$ non-diagonal generators are defined as
\begin{equation}
\label{eq:fundamental_non-diagonal_generators}
\left(T_{\mbox{\tiny{Fund}}}^{(i,j;1)}\right)_{a,b} = \frac{1}{2} \left( \delta_{a,i}\delta_{b,j} + \delta_{a,j}\delta_{b,i}\right), \qquad \left(T_{\mbox{\tiny{Fund}}}^{(i,j;2)}\right)_{a,b} = \frac{1}{2i} \left( \delta_{a,i}\delta_{b,j} - \delta_{a,j}\delta_{b,i}\right),
\end{equation}
for $1 \le i < j \le N$. Note that, with these conventions, for $N=2$ the generators in the fundamental representation reduce to $\frac{1}{2}\sigma^a$ ($\sigma^a$ denoting a Pauli matrix), while for $N=3$ the generators in the fundamental representation are $\frac{1}{2}\lambda^a$ (where $\lambda^a$ denotes a Gell-Mann matrix).

Given a generic element $u$ of the $SU(N)$ group in the fundamental representation, the same group element in the 2AS representation is a complex-valued matrix of size $(N(N-1)/2) \times (N(N-1)/2)$ whose entries (which can be labelled by pairs of indices $(i,j)$, with $1 \le i < j \le N$) are defined as
\begin{align}
\label{2as}
U_{(i,j) (k,l)} = -2 \Tr \left( T_{\mbox{\tiny{Fund}}}^{(i,j;2)\, {\tiny{T}}}\, u\, T_{\mbox{\tiny{Fund}}}^{(k,l;2)}\, u^{\tiny{T}} \right).
\end{align}
Interpreting Eq.~(\ref{2as}) as a map $R$ from the group elements in the fundamental representation to group elements in the two-index antisymmetric representation, the generators in the latter representation can then be obtained from the pushforward $R_\star$ of the generators in the fundamental representation. Equivalently, if one considers group elements of the form $u=\ide_N + i \epsilon T_{\mbox{\tiny{fund}}}^a + O(\epsilon^2)$, the corresponding generator in the 2AS representation can be obtained as $\lim_{\epsilon \to 0} (-i/\epsilon)\left(U - \ide_{N(N-1)/2}\right)$.

Explicitly, with these conventions the $SU(4)$ generators in the fundamental representation take the form:
\begin{small}
\begin{align}
\label{SU4_fund}
& T^1_{\mbox{\tiny{Fund}}} = \frac{1}{2}\left(
\begin{array}{cccc}
0 & 1 & 0 & 0 \\
1 & 0 & 0 & 0 \\
0 & 0 & 0 & 0 \\
0 & 0 & 0 & 0
\end{array}
\right), \,\,
T^2_{\mbox{\tiny{fund}}} = \frac{1}{2}\left(
\begin{array}{cccc}
0 & -i & 0 & 0 \\
i & 0 & 0 & 0 \\
0 & 0 & 0 & 0 \\
0 & 0 & 0 & 0
\end{array}
\right), \,\,
T^3_{\mbox{\tiny{Fund}}} = \frac{1}{2} \left(
\begin{array}{ccccc}
1 & 0 & 0 & 0 \\
0 & -1 & 0 & 0 \\
0 & 0 & 0 & 0 \\
0 & 0 & 0 & 0
\end{array}
\right), \nonumber \\
& T^4_{\mbox{\tiny{Fund}}} = \frac{1}{2} \left(
\begin{array}{cccc}
0 & 0 & 1 & 0 \\
0 & 0 & 0 & 0 \\
1 & 0 & 0 & 0 \\
0 & 0 & 0 & 0
\end{array}
\right), \,\,
T^5_{\mbox{\tiny{Fund}}} = \frac{1}{2} \left(
\begin{array}{cccc}
0 & 0 & -i & 0 \\
0 & 0 & 0 & 0 \\
i & 0 & 0 & 0 \\
0 & 0 & 0 & 0
\end{array}
\right), \,\,
T^6_{\mbox{\tiny{fund}}} = \frac{1}{2}\left(
\begin{array}{cccc}
0 & 0 & 0 & 0 \\
0 & 0 & 1 & 0 \\
0 & 1 & 0 & 0 \\
0 & 0 & 0 & 0
\end{array}
\right), \nonumber \\
& T^7_{\mbox{\tiny{Fund}}} = \frac{1}{2}\left(
\begin{array}{cccc}
0 & 0 & 0 & 0 \\
0 & 0 & -i & 0 \\
0 & i & 0 & 0 \\
0 & 0 & 0 & 0
\end{array}
\right), \,\,
T^8_{\mbox{\tiny{Fund}}} = \frac{1}{2\sqrt{3}} \left(
\begin{array}{cccc}
1 & 0 & 0 & 0 \\
0 & 1 & 0 & 0 \\
0 & 0 & -2 & 0 \\
0 & 0 & 0 & 0
\end{array}
\right), \,\,
T^9_{\mbox{\tiny{Fund}}} = \frac{1}{2} \left(
\begin{array}{cccc}
0 & 0 & 0 & 1 \\
0 & 0 & 0 & 0 \\
0 & 0 & 0 & 0 \\
1 & 0 & 0 & 0
\end{array}
\right), \nonumber \\
& T^{10}_{\mbox{\tiny{Fund}}} = \frac{1}{2} \left(
\begin{array}{cccc}
0 & 0 & 0 & -i \\
0 & 0 & 0 & 0 \\
0 & 0 & 0 & 0 \\
i & 0 & 0 & 0
\end{array}
\right), \,\,
T^{11}_{\mbox{\tiny{fund}}} = \frac{1}{2} \left(
\begin{array}{cccc}
0 & 0 & 0 & 0 \\
0 & 0 & 0 & 1 \\
0 & 0 & 0 & 0 \\
0 & 1 & 0 & 0
\end{array}
\right), \,\,
T^{12}_{\mbox{\tiny{fund}}} = \frac{1}{2} \left(
\begin{array}{cccc}
0 & 0 & 0 & 0 \\
0 & 0 & 0 & -i \\
0 & 0 & 0 & 0 \\
0 & i & 0 & 0
\end{array}
\right), \nonumber \\
& T^{13}_{\mbox{\tiny{Fund}}} = \frac{1}{2} \left(
\begin{array}{cccc}
0 & 0 & 0 & 0 \\
0 & 0 & 0 & 0 \\
0 & 0 & 0 & 1 \\
0 & 0 & 1 & 0
\end{array}
\right), \,\,
T^{14}_{\mbox{\tiny{Fund}}} = \frac{1}{2} \left(
\begin{array}{cccc}
0 & 0 & 0 & 0 \\
0 & 0 & 0 & 0 \\
0 & 0 & 0 & -i \\
0 & 0 & i & 0
\end{array}
\right), \,\,
T^{15}_{\mbox{\tiny{Fund}}} = \frac{1}{2\sqrt{6}} \left(
\begin{array}{cccc}
1 & 0 & 0 & 0 \\
0 & 1 & 0 & 0 \\
0 & 0 & 1 & 0 \\
0 & 0 & 0 & -3
\end{array}
\right), \qquad
\end{align}
\end{small}
while the $SU(4)$ generators in the two-index antisymmetric representation read:

\begin{footnotesize}
\begin{align}
\label{SU4_2a}
& \hspace{-20mm} T^1_{\mbox{\tiny{2AS}}} = \frac{1}{2}\left(
\begin{array}{cccccc}
0 & 0 & 0 & 0 & 0 & 0 \\
0 & 0 & 1 & 0 & 0 & 0 \\
0 & 1 & 0 & 0 & 0 & 0 \\
0 & 0 & 0 & 0 & 1 & 0 \\
0 & 0 & 0 & 1 & 0 & 0 \\
0 & 0 & 0 & 0 & 0 & 0
\end{array}
\right), \,\,\,\,\,\,\,\,\,\,\,
T^2_{\mbox{\tiny{2AS}}} = \frac{1}{2}\left(
\begin{array}{cccccc}
0 & 0 & 0 & 0 & 0 & 0 \\
0 & 0 & -i & 0 & 0 & 0 \\
0 & i & 0 & 0 & 0 & 0 \\
0 & 0 & 0 & 0 & -i & 0 \\
0 & 0 & 0 & i & 0 & 0 \\
0 & 0 & 0 & 0 & 0 & 0
\end{array}
\right), \,\,\,\,\,\,\,\,\,\,\,\,\,\,
T^3_{\mbox{\tiny{2AS}}} = \frac{1}{2}\left(
\begin{array}{cccccc}
0 & 0 & 0 & 0 & 0 & 0 \\
0 & 1 & 0 & 0 & 0 & 0 \\
0 & 0 & -1 & 0 & 0 & 0 \\
0 & 0 & 0 & 1 & 0 & 0 \\
0 & 0 & 0 & 0 & -1 & 0 \\
0 & 0 & 0 & 0 & 0 & 0
\end{array}
\right), \nonumber \\
& \hspace{-20mm} T^4_{\mbox{\tiny{2AS}}} = \frac{1}{2}\left(
\begin{array}{cccccc}
0 & 0 & -1 & 0 & 0 & 0 \\
0 & 0 & 0 & 0 & 0 & 0 \\
-1 & 0 & 0 & 0 & 0 & 0 \\
0 & 0 & 0 & 0 & 0 & 1 \\
0 & 0 & 0 & 0 & 0 & 0 \\
0 & 0 & 0 & 1 & 0 & 0
\end{array}
\right), \,\,
T^5_{\mbox{\tiny{2AS}}} = \frac{1}{2}\left(
\begin{array}{cccccc}
0 & 0 & i & 0 & 0 & 0 \\
0 & 0 & 0 & 0 & 0 & 0 \\
-i & 0 & 0 & 0 & 0 & 0 \\
0 & 0 & 0 & 0 & 0 & -i \\
0 & 0 & 0 & 0 & 0 & 0 \\
0 & 0 & 0 & i & 0 & 0
\end{array}
\right), \,\,\,\,\,\,\,\,\,\,\,\,\,\,
T^6_{\mbox{\tiny{2AS}}} = \frac{1}{2}\left(
\begin{array}{cccccc}
0 & 1 & 0 & 0 & 0 & 0 \\
1 & 0 & 0 & 0 & 0 & 0 \\
0 & 0 & 0 & 0 & 0 & 0 \\
0 & 0 & 0 & 0 & 0 & 0 \\
0 & 0 & 0 & 0 & 0 & 1 \\
0 & 0 & 0 & 0 & 1 & 0
\end{array}
\right), \nonumber \\
& \hspace{-20mm} T^7_{\mbox{\tiny{2AS}}} = \frac{1}{2}\left(
\begin{array}{cccccc}
0 & -i & 0 & 0 & 0 & 0 \\
i & 0 & 0 & 0 & 0 & 0 \\
0 & 0 & 0 & 0 & 0 & 0 \\
0 & 0 & 0 & 0 & 0 & 0 \\
0 & 0 & 0 & 0 & 0 & -i \\
0 & 0 & 0 & 0 & i & 0
\end{array}
\right), \,\,
T^8_{\mbox{\tiny{2AS}}} = \frac{1}{\sqrt{12}}\left(
\begin{array}{cccccc}
2 & 0 & 0 & 0 & 0 & 0 \\
0 & -1 & 0 & 0 & 0 & 0 \\
0 & 0 & -1 & 0 & 0 & 0 \\
0 & 0 & 0 & 1 & 0 & 0 \\
0 & 0 & 0 & 0 & 1 & 0 \\
0 & 0 & 0 & 0 & 0 & -2
\end{array}
\right), \,\,
T^9_{\mbox{\tiny{2AS}}} = \frac{1}{2}\left(
\begin{array}{cccccc}
0 & 0 & 0 & 0 & -1 & 0 \\
0 & 0 & 0 & 0 & 0 & -1 \\
0 & 0 & 0 & 0 & 0 & 0 \\
0 & 0 & 0 & 0 & 0 & 0 \\
-1 & 0 & 0 & 0 & 0 & 0 \\
0 & -1 & 0 & 0 & 0 & 0
\end{array}
\right), \nonumber \\
& \hspace{-20mm} T^{10}_{\mbox{\tiny{2AS}}} = \frac{1}{2}\left(
\begin{array}{cccccc}
0 & 0 & 0 & 0 & i & 0 \\
0 & 0 & 0 & 0 & 0 & i \\
0 & 0 & 0 & 0 & 0 & 0 \\
0 & 0 & 0 & 0 & 0 & 0 \\
-i & 0 & 0 & 0 & 0 & 0 \\
0 & -i & 0 & 0 & 0 & 0
\end{array}
\right), \,\,
T^{11}_{\mbox{\tiny{2AS}}} = \frac{1}{2}\left(
\begin{array}{cccccc}
0 & 0 & 0 & 1 & 0 & 0 \\
0 & 0 & 0 & 0 & 0 & 0 \\
0 & 0 & 0 & 0 & 0 & -1 \\
1 & 0 & 0 & 0 & 0 & 0 \\
0 & 0 & 0 & 0 & 0 & 0 \\
0 & 0 & -1 & 0 & 0 & 0
\end{array}
\right), \,\,\,\,\,\,\,\,
T^{12}_{\mbox{\tiny{2AS}}} = \frac{1}{2}\left(
\begin{array}{cccccc}
0 & 0 & 0 & -i & 0 & 0 \\
0 & 0 & 0 & 0 & 0 & 0 \\
0 & 0 & 0 & 0 & 0 & i \\
i & 0 & 0 & 0 & 0 & 0 \\
0 & 0 & 0 & 0 & 0 & 0 \\
0 & 0 & -i & 0 & 0 & 0
\end{array}
\right), \nonumber \\
& \hspace{-20mm} T^{13}_{\mbox{\tiny{2AS}}} = \frac{1}{2}\left(
\begin{array}{cccccc}
0 & 0 & 0 & 0 & 0 & 0 \\
0 & 0 & 0 & 1 & 0 & 0 \\
0 & 0 & 0 & 0 & 1 & 0 \\
0 & 1 & 0 & 0 & 0 & 0 \\
0 & 0 & 1 & 0 & 0 & 0 \\
0 & 0 & 0 & 0 & 0 & 0
\end{array}
\right), \,\,\,\,\,\,\,\,\,
T^{14}_{\mbox{\tiny{2AS}}} = \frac{1}{2}\left(
\begin{array}{cccccc}
0 & 0 & 0 & 0 & 0 & 0 \\
0 & 0 & 0 & -i & 0 & 0 \\
0 & 0 & 0 & 0 & -i & 0 \\
0 & i & 0 & 0 & 0 & 0 \\
0 & 0 & i & 0 & 0 & 0 \\
0 & 0 & 0 & 0 & 0 & 0
\end{array}
\right), \,\,\,\,\,\,\,
T^{15}_{\mbox{\tiny{2AS}}} = \frac{1}{\sqrt{6}}\left(
\begin{array}{cccccc}
1 & 0 & 0 & 0 & 0 & 0 \\
0 & 1 & 0 & 0 & 0 & 0 \\
0 & 0 & 1 & 0 & 0 & 0 \\
0 & 0 & 0 & -1 & 0 & 0 \\
0 & 0 & 0 & 0 & -1 & 0 \\
0 & 0 & 0 & 0 & 0 & -1
\end{array}
\right). \nonumber \\
\end{align}
\end{footnotesize}

\section{Autocorrelation Times}
\label{sec:autocorrelation}

In an HMC algorithm, gauge configurations are generated through a Markov chain process. As a consequence, subsequent configurations can be correlated. Accounting for autocorrelation is essential in the estimation of observables through the Monte Carlo average. Consider a succession $a_i$ of measurements of an observable $A$. We estimate the integrated autocorrelation time $\tau_{int}$ of the observable $A$ as
\begin{equation}\label{eq:def_tauint}
    \tau_{int} = \oneh + \sum_{t=1}^{W} \frac{\Gamma(t)}{\Gamma(0)} \; ,
\end{equation}
where $t$ is the Monte Carlo time, $W$ is the summation window, and $\Gamma(t)$ is our approximation of the autocorrelation function of the series:
\begin{equation}
    \Gamma(t) \simeq \frac{1}{N-t} \sum_{n=1}^{N-t} \left( a_n - \braket{a_-} \right) \left( a_{n+t} - \braket{a_+} \right) \; , \;\;\;\;\; 0 \leq t < N \; .
\end{equation}
$\braket{a_-}$ and $\braket{a_+}$ are the average over $N-t$ measurements, respectively the first and the last ones. In order to choose a summation window, we check at which Monte Carlo time $t$ the function $\Gamma(t)/\Gamma(0)$ is smaller than twice its variance \cite{DelDebbio:2007pz}, which is estimated within the Madras Sokal approximation \cite{Madras:1988ei}.
\begin{figure}[h!]
    \centering
    \includegraphics[width=0.45\textwidth]{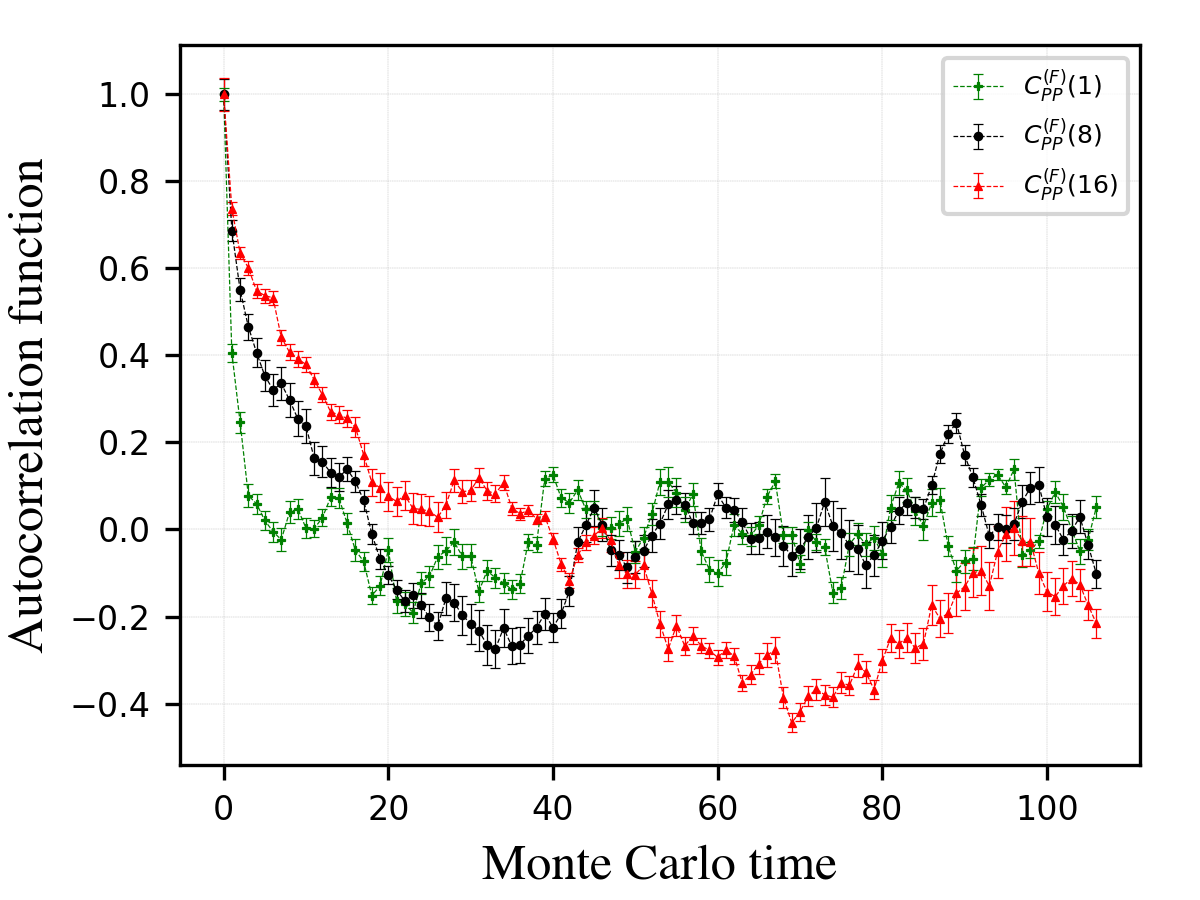}
    \includegraphics[width=0.45\textwidth]{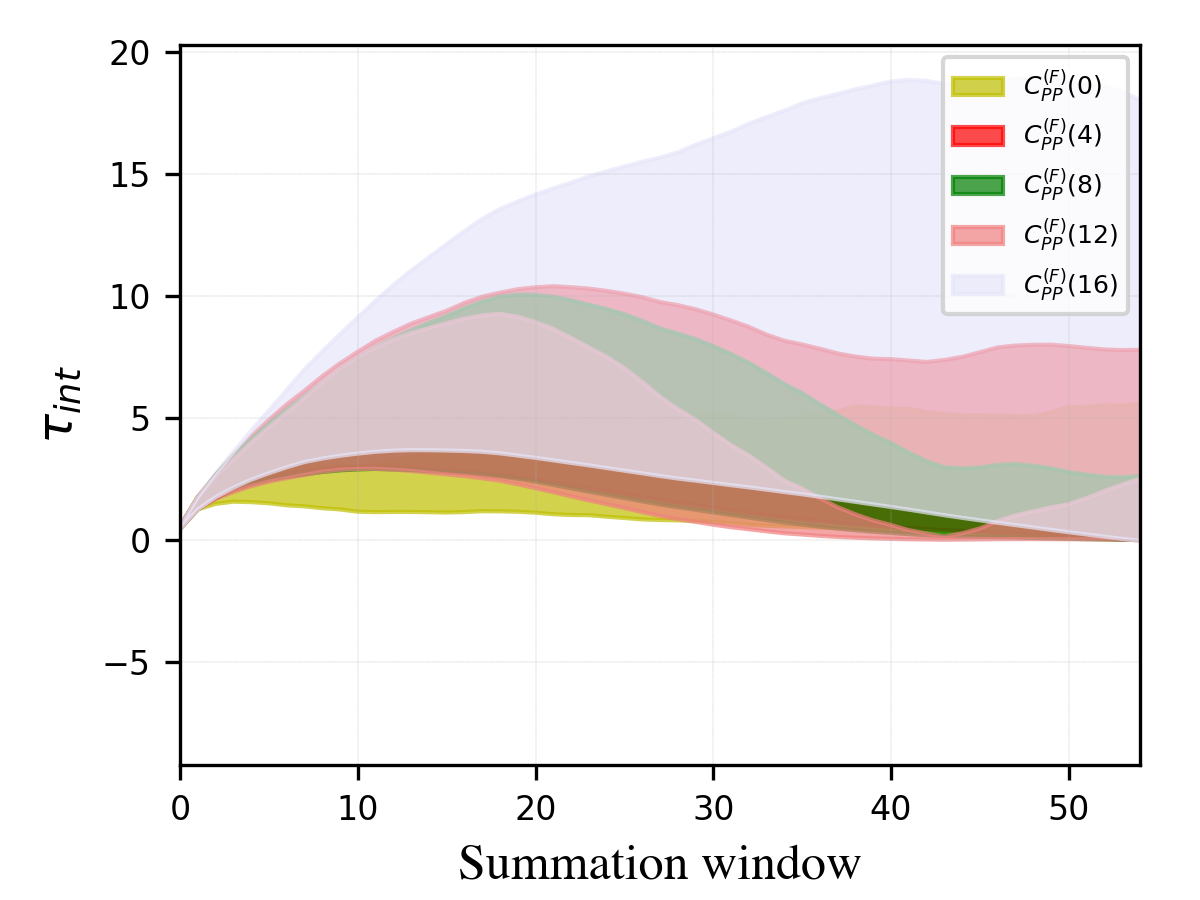}
    \caption{Left panel: autocorrelation function $\Gamma(t)/\Gamma(0)$ compute for the two point function of two pseudoscalar mesons with fermions in the fundamental representation. Different colors represent different intervals in lattice time between the source and the sink. Right panel: integrated autocorrelation time as a function of the summation window, computed for the same correlator on the left side at several times. The two point function is obtained from the ensemble A0.}
    \label{fig:autocorrelation}
\end{figure}
Typical plots that we obtain for these quantities are shown in Fig. \ref{fig:autocorrelation} for the pseudoscalar-pseudoscalar correlator of fundamental fermions at different lattice times. In the left panel, our estimation of the autocorrelation function $\Gamma(t)/\Gamma(0)$ is plotted as a function of Monte Carlo time, showing the typical decaying behaviour. The right panel shows the integrated autocorrelation time $\tau_{int}$ of Eq.~\eqref{eq:def_tauint} for an increasing summation window $W$. In both panels, the correlation functions are computed for operators at different lattice times in order to monitor both the short and long distance behaviour. \\
By knowing the correlation of an observable between trajectories in the HMC, we can establish how many of them have to be skipped performing the measurement. This is our strategy for the computation of the correlators on our ensembles. Alternatively, the naive estimate for the statistical error $\sigma_0$ of a given observable can be corrected to be $\sigma^2=2\tau_{int} \sigma_0^2$. \\
Accounting for autocorrelation has been essential in the following analysis, especially for those quantities computed in the proximity of the chiral limit, where the autocorrelation was more significant.

 \section{Measurements of Correlation Functions}
 \label{app:sec:measure}
 
In this section we discuss the details behind the measurement of correlation functions $C_{ab}^{R}(t) \equiv C_{ab}^{R}(t,0)$
\begin{equation}\label{eq:app:generic_lattice_correlator_2pt}
    C_{ab}^{R}(t,0) = \frac{1}{L^3} \sum_{\vet{x}} \braket{O^{R}_a(\vet{x},t) \bar{O}^{R}_b(0,t_i)} \; , \;\;\;\;\; a,b = P,A,V \; ,
\end{equation}
where $R$ denotes the representation of the fermionic fields. For each gauge configuration $n=1,\dots N_{cfg}$ we average measurements performed every $4$ source-times $t_i$, resulting in the correlator $C_{ab, \, n}^{R}(t)$. 
\begin{equation}
C_{ab, \, n}^{R} (t) = \frac{8}{T}\sum_{t_i=0,4,\dots}^{T/2} C_{ab\, n}^{F}(t,t_i)
\end{equation}
The gauge average $C_{PP}^{R}(t)$ and its variance $\delta C_{ab}^{R}(t)^2$ are then estimated from a new set of $N\p$ correlators, $C_{ab, \, n\p}^{R}$, resampled with a bootstrap procedure from $C_{ab, \, n}^{R}$. In this step, trajectories are discarded in order to account for the autocorrelation that is monitored across observables and ensembles according to Appendix \ref{sec:autocorrelation}
\begin{equation}
    C_{ab}^{R}(t) = \frac{1}{N\p} \sum_{n\p=0}^{N\p-1} C_{ab, \, n\p}^{R} \; , \;\;\;\;\; \delta C_{ab}^{R}(t)^2 =  \dfrac{1}{N\p} \sum_{n\p=0}^{N\p-1} \left[ C_{ab, \, n\p}^{R} -  C_{ab}^{R}(t) \right]^2
\end{equation} 
In order to suppress the excited states created by the hadronic operator in Eq.~\eqref{eq:app:generic_lattice_correlator_2pt}, we compute the two point functions by using different types of smearing on the fermionic fields. The procedure is not gauge invariant and requires working at fixed gauge. In particular, we adopt local ($\psi^{R}(x)$) and smeared ($\tilde{\psi}^{R}(x)$) fields
\begin{equation}
\begin{split}
    & \psi^{R}(x)_\alpha^c = \int dy\, \delta(x-y) \delta_{\alpha \alpha\p} \delta_{cc\p} \, \psi^{R}(y)_{\alpha\p}^{c\p} 
    ; ,\\
    & \tilde{\psi}_g^{R}(x)_\alpha^c =  \int dy\, \frac{e^{-(x-y)^2/2g^2}}{\sqrt{2\pi}g} \delta_{\alpha \alpha\p} \delta_{cc\p}  \psi^R(y)_{\alpha\p}^{c\p} \; ,
\end{split}
\end{equation}
where $c, c'$ are color indices, $\alpha, \alpha\p$ are Dirac indices and a sum is intended over $\alpha\p, c\p$. The parameter $g$ can is tuned according to Section \ref{sec::measurements} in order to suppress the excited states in the spectral reconstruction. By combining local and smeared operators, we obtain three types of correlators: local-local, smeared-smeared, local-smeared. In the latter, the operators at the source and the sink are different, and this can produce negative contributions to the spectral density. As demonstrated in the Appendix of \cite{Hansen_2019}, such terms do not jeopardise the extraction of the spectral density from the correlators.
\newpage

\section{Ensembles}
\label{app:ensembles}

\begin{table}[h!]
    \centering
    \begin{tabular}{c|c c c c c c c}
          & $am_0^F$ & $am_0^{2AS}$ & $\braket{P}$ & $N_{cfg}$  & $am_{PCAC}^F$ & $am_{PCAC}^{2AS}$\\[-1em]\\ \hline
          \\[-0.8em]
          A0 & -0.45 & -0.45 & 0.60893(2) & 216 & 0.0468(23) & 0.2327(37) \\
          \\[-0.8em]
          A1 & -0.455 & -0.45 & 0.60896(3) & 99 & 0.0386(17) & 0.2311(13)\\
          \\[-0.8em]
          A2 &  -0.46 & -0.45 & 0.61392(2) & 694 & 0.0332(26) & 0.2290(30) \\ 
          \\[-0.8em]
          A3 & -0.465 & -0.45 & 0.60917(4) & 82  & 0.0262(14) & 0.2280(19) \\
          \\[-0.8em]
          A4 & -0.47 & -0.45 & 0.60942(1) & 446   & 0.0209(29) & 0.2277(35) \\
          \\[-0.8em]
          B0 & -0.45 & -0.54 & 0.61181(4) & 65 & 0.0321(78) & 0.1200(53) \\
         \\[-0.8em]
         B1 &-0.45 & -0.56 & 0.61259(2) & 232   & 0.0338(94) & 0.0988(40)\\
         \\[-0.8em]
         B2 & -0.45 & -0.58 & 0.61344(4) & 243 & 0.0313(16) & 0.0691(35)\\
         \\[-0.8em]
         B3 & -0.45 & -0.59 & 0.61392(2)  & 180  & 0.0306(36) & 0.0544(19) \\
         \\[-0.8em] 
         B3 & -0.45 & -0.60 &0.61427(2) & 176 & 0.0480(17) & 0.0292(12) \\
         \\[-0.8em]
         C0 & -0.47 & -0.48 & 0.61011(3)  & 116 & 0.0205(34)& 0.1962(54)\\
         \\[-0.8em]
         C1 & -0.47 & -0.52 & 0.61148(3)& 78 & 0.0170(67) & 0.1415(41)\\
         \\[-0.8em]
         C2 & -0.47 & -0.53 & 0.61203(3)& 90  &0.0143(56) & 0.1279(39)\\
         \\[-0.8em]
         C3 & -0.47  & -0.54 &0.61248(3) & 92  &0.0114(45) & 0.1126(32)\\
         \\[-0.8em]
         C4 & -0.47 &-0.58 & 0.61398(6) & 53   & 0.0068(15) & 0.0651(27)\\
         \\[-0.8em]
         S0 & -0.44 & -0.60 & 0.61403(5) & 43 & 0.0454(10) & 0.0495(20) \\
         \\[-0.8em]
    \end{tabular}
    \caption{Ensembles used to extrapolate the chiral limit of the $SU(4)$ gauge theory with two fundamental and two  two-index antisymmetric fermions. The coupling is $\beta=11$, and the volume of the lattice is $16^3 \times 32$.}
    \label{tab:ensembles}
\end{table}
\newpage

\section{Masses}
\label{app:sec:masses}

\begin{table}[h]
     \centering
     \begin{tabular}{c c c}
          $am_0^{(F)}$ &  $am_{PCAC}^{(F)}$ & $am_{PCAC}^{(2AS)}$ \\[-1em]\\ \hline
          \\[-0.8em]
           -0.45 & 0.0468(23) & 0.2327(37) \\
          \\[-0.8em]
          -0.455 & 0.0386(17) &  0.2311(13) \\
          \\[-0.8em]
          -0.46 & 0.0332(26) & 0.2290(30)\\
          \\[-0.8em]
          -0.465 & 0.0262(14) &  0.2280(19) \\
          \\[-0.8em]
          -0.47 &  0.0209(29)&  0.2277(35) \\
          [-0.8em]
         \\ \hline
     \end{tabular}
     \caption{PCAC masses for the fundamental representation used in the chiral extrapolation. The bare mass of the antisymmetric fermions is fixed at $am_0^{(2AS)}=-0.45$. They correspond to the ensembles A0-A4.}
     \label{tab:pcac_fund}
 \end{table}
 
  \begin{table}[h]
     \centering
     \begin{tabular}{c c c | c c c}
          $am_0^{(2AS)}$ &  $am_{PCAC}^{(2AS)}$ &  $am_{PCAC}^{(F)}$ & $am_0^{(2AS)}$ &  $am_{PCAC}^{(2AS)}$ &  $am_{PCAC}^{(F)}$\\[-1em]\\ \hline
          \\[-0.8em]
           -0.54 & 0.1200(53) & 0.0321(78) &  -0.48 & 0.1962(54) & 0.0205(34)  \\
          \\[-0.8em]
          -0.56 & 0.0988(40) & 0.0338(94) & -0.52 & 0.1415(41) & 0.0170(67)\\
          \\[-0.8em]
          -0.58 & 0.0691(35) & 0.0313(16) & -0.53 & 0.1279(39) & 0.0143(56)  \\
          \\[-0.8em]
          -0.59 & 0.0544(19) & 0.0306(36) & -0.54 & 0.1126(32) & 0.0114(45)\\
          \\[-0.8em]
          -0.60 &  0.0480(17) & 0.0292(12)&  -0.58 &  0.0651(27) & 0.0068(15) \\
          [-0.8em]
         \\ \hline
     \end{tabular}
     \caption{PCAC masses for the antisymmetric representation used in the chiral extrapolation. The values on the left are obtained with $am_0^{(F)}=-0.45$ (ensembles B0-B4), the ones on the right with $am_0^{(F)}=-0.47$ (ensembles C0-C4).}
     \label{tab:pcac_2as}
 \end{table}
 \begin{table}[h]
     \centering
     \begin{tabular}{c c c | c c c}
          $am_0^{(F)}$ &  $aM_{PP}^{(F)}$ & $aM_{PP}^{(2AS)}$ & $am_0^{(2AS)}$ &  $aM_{PP}^{(2AS)}$ & $aM_{PP}^{(F)}$ \\[-1em]\\ \hline
          \\[-0.8em]
           -0.45 & 0.3555(38) &  0.8037(14) & -0.54  & 0.5426(22) &0.375(36) \\ 
          \\[-0.8em]
          -0.455 & 0.290(60) &  0.7701(31) &-0.56 & 0.4638(33) & 0.3114(69)  \\
          \\[-0.8em]
          -0.46 & 0.280(16)  & 0.7640(28) & -0.58 & 0.4035(24) & 0.3147(40) \\
          \\[-0.8em]
          -0.465 & 0.254(16)  & 0.7635(29)  & -0.59 & 0.3600(38) & 0.3038(67)\\
          \\[-0.8em]
          -0.47 & 0.255(10)  &  0.7634(20) &-0.60  & 0.3407(48) & 0.321(13) \\
          [-0.8em]
         \\ \hline
     \end{tabular}
     \caption{Masses of the pseudoscalar mesons. On the left, we vary the bare fundamental mass and keeping $am_0(2AS)=-0.45$ (ensembles A0-A4). Conversely, the right the fundamental bare mass is fixed at $am_0(Fund)=-0.45$ (ensembles B0-B4).}
     \label{tab:PSmesons}
 \end{table}
\begin{table}[h]
    \centering
    \begin{tabular}{c c c c}
        $ a \sigma$ & $a M_{\mathrm{PP}}^{\mathrm{(2AS)}}$ & $\chi^2_{f_\sigma^{(n)}}$ / d.o.f. & $n$ \\[-1em]\\ \hline
          \\[-0.8em]
         0.18 & 0.3558(30) & 0.61 & 2\\
         0.19 & 0.3505(33) & 0.75  & 2\\
         0.2 & 0.3550(23) & 1.36  & 2\\
         0.21 & 0.3554(22) & 0.81 & 2\\
         0.22 & 0.3523(26) & 1.18 & 2\\
         0.23 & 0.3580(22) & 2.13 & 2\\
         0.24 & 0.3557(24) & 1.58 & 2\\
         0.19 & 0.3498(37) & 0.71 & 3\\
         0.2 & 0.3516(32) & 0.96  & 3\\
         0.21 & 0.3534(29) & 1.80 & 3\\
         0.22 & 0.3505(33) & 1.38 & 3\\
         0.23 & 0.3592(25) & 2.04 & 3\\
    \end{tabular}
    \caption{Fit results for $a M_{PP}^{(2AS)}$ from smeared spectral densities for different smearing radii $\sigma$ and different number of states $n$. These values appear in Fig. \ref{fig:finalres}.}
    \label{tab:twogau_e0}
\end{table}
\newpage

\section{Isospin Generators}
\label{sec:app:iso}
The isospin group of the 2AS sector in the 2-flavor Ferretti model is $SO(4)$. The Goldstone bosons arising from the $SU(4)/SO(4)$ cosets transform in a 9-dimensional representation of the isospin, whose generators $X_n$, $n=1,\dots 6$ are listed in this appendix. These are obtained according to the convention of \cite{Cossu_2019} regarding the generators and the structure constants of $SU(4)$.
\begin{equation}
    X_1 = -\frac{i}{2} \smatrix    0 & -2 & 0 & 0 & 0 & 0 & 0 & 0 & 0 \\
                                         2 & 0 & 0 & 0 & 0 & 0 & 0 & 0 & 0 \\
                                         0 & 0 & 0 & 1 & 0 & 0 & 0 & 0 & 0 \\
                                         0 & 0 & -1 & 0 & 0 & 0 & 0 & 0 & 0 \\
                                         0 & 0 & 0 & 0 & 0 & 0 & 0 & 0 & 0 \\
                                         0 & 0 & 0 & 0 & 0 & 0 & 1 & 0 & 0 \\
                                         0 & 0 & 0 & 0 & 0 & -1 & 0 & 0 & 0 \\
                                         0 & 0 & 0 & 0 & 0 & 0 & 0 & 0 & 0 \\
                                         0 & 0 & 0 & 0 & 0 & 0 & 0 & 0 & 0 \cmatrix \; ,
\end{equation}

\begin{equation}
    X_2 = -\frac{i}{2} \smatrix  0 & 0 & 0 & 1 & 0 & 0 & 0 & 0 & 0 \\
                                         0 & 0 & 1 & 0 & 0 & 0 & 0 & 0 & 0 \\
                                         0 & -1 & 0 & 0 & -\sqrt{3} & 0 & 0 & 0 & 0 \\
                                         -1 & 0 & 0 & 0 & 0 & 0 & 0 & 0 & 0 \\
                                         0 & 0 & \sqrt{3} & 0 & 0 & 0 & 0 & 0 & 0 \\
                                         0 & 0 & 0 & 0 & 0 & 0 & 0 & 1 & 0 \\
                                         0 & 0 & 0 & 0 & 0 & 0 & 0 & 0 & 0 \\
                                         0 & 0 & 0 & 0 & 0 & -1 & 0 & 0 & 0 \\
                                         0 & 0 & 0 & 0 & 0 & 0 & 0 & 0 & 0 \cmatrix \; ,
\end{equation}
\begin{equation}
    X_3 = -\frac{i}{2}\smatrix       0 & 0 & 1 & 0 & 0 & 0 & 0 & 0 & 0 \\
                                           0 & 0 & 0 & -1 & 0 & 0 & 0 & 0 & 0 \\
                                           -1 & 0 & 0 & 0 & 0 & 0 & 0 & 0 & 0 \\
                                           0 & 1 & 0 & 0 & -\sqrt{3} & 0 & 0 & 0 & 0 \\
                                           0 & 0 & 0 & \sqrt{3} & 0 & 0 & 0 & 0 & 0 \\
                                           0 & 0 & 0 & 0 & 0 & 0 & 0 & 0 & 0 \\
                                           0 & 0 & 0 & 0 & 0 & 0 & 0 & 1 & 0 \\
                                           0 & 0 & 0 & 0 & 0 & 0 & -1 & 0 & 0 \\
                                           0 & 0 & 0 & 0 & 0 & 0 & 0 & 0 & 0 \cmatrix \; ,
\end{equation}

\begin{equation}
    X_4 = -\frac{i}{2}\smatrix     0 & 0 & 0 & 0 & 0 & 0 & 1 & 0 & 0 \\
                                           0 & 0 & 0 & 0 & 0 & 1 & 0 & 0 & 0 \\
                                           0 & 0 & 0 & 0 & 0 & 0 & 0 & 1 & 0 \\
                                           0 & 0 & 0 & 0 & 0 & 0 & 0 & 0 & 0 \\
                                           0 & 0 & 0 & 0 & 0 & 1/\sqrt{3} & 0 & 0 & 0 \\
                                           0 & -1 & 0 & 0 & -1/\sqrt{3} & 0 & 0 & 0 & -2\sqrt{2/3} \\
                                           -1 & 0 & 0 & 0 & 0 & 0 & 0 & 0 & 0 \\
                                           0 & 0 & -1 & 0 & 0 & 0 & 0 & 0 & 0 \\
                                           0 & 0 & 0 & 0 & 0 & 2\sqrt{2/3} & 0 & 0 & 0 \cmatrix \; ,
\end{equation}

\begin{equation}
    X_5 =-\frac{i}{2}\smatrix      0 & 0 & 0 & 0 & 0 & 1 & 0 & 0 & 0 \\
                                           0 & 0 & 0 & 0 & 0 & 0 & -1 & 0 & 0 \\
                                           0 & 0 & 0 & 0 & 0 & 0 & 0 & 0 & 0 \\
                                           0 & 0 & 0 & 0 & 0 & 0 & 0 & 1 & 0 \\
                                           0 & 0 & 0 & 0 & 0 & 0 & 1/\sqrt{3} & 0 & 0 \\
                                           -1 & 0 & 0 & 0 & 0 & 0 & 0 & 0 & 0 \\
                                           0 & 1 & 0 & 0 & -1/\sqrt{3} & 0 & 0 & 0 & -2\sqrt{2/3} \\
                                           0 & 0 & 0 & -1 & 0 & 0 & 0 & 0 & 0 \\
                                           0 & 0 & 0 & 0 & 0 & 0 & 2\sqrt{2/3} & 0 & 0 \cmatrix\; , 
\end{equation}
\begin{equation}
    X_6 = -\frac{i}{2} \smatrix      0 & 0 & 0 & 0 & 0 & 0 & 0 & 0 & 0 \\
                                           0 & 0 & 0 & 0 & 0 & 0 & 0 & 0 & 0 \\
                                           0 & 0 & 0 & 0 & 0 & 1 & 0 & 0 & 0 \\
                                           0 & 0 & 0 & 0 & 0 & 0 & 1 & 0 & 0 \\
                                           0 & 0 & 0 & 0 & 0 & 0 & 0 & -2/\sqrt{3} & 0 \\
                                           0 & 0 & -1 & 0 & 0 & 0 & 0 & 0 & 0 \\
                                           0 & 0 & 0 & -1 & 0 & 0 & 0 & 0 & 0 \\
                                           0 & 0 & 0 & 0 & 2/\sqrt{3} & 0 & 0 & 0 & -2\sqrt{2/3} \\
                                           0 & 0 & 0 & 0 & 0 & 0 & 0 & 2\sqrt{2/3} & 0 \cmatrix
\end{equation}

\bibliographystyle{unsrt}         
\bibliography{main.bib}
\end{document}